\documentclass[lettersize,journal]{IEEEtran}
\usepackage[backend=bibtex,bibstyle=ieee,citestyle=numeric-comp,doi=false]{biblatex}
\usepackage{amsmath,amsfonts,amssymb}
\usepackage{algorithmic}
\usepackage{algorithm}
\usepackage{array}
\usepackage[caption=false,font=normalsize,labelfont=sf,textfont=sf]{subfig}
\usepackage{textcomp}
\usepackage{stfloats}
\usepackage{url}
\usepackage{verbatim}
\usepackage{graphicx}
\usepackage{circuitikz}
\usepackage{cleveref}
\usepackage{multirow,multicol}
\crefrangeformat{equation}{equations~(#3#1#4) to~(#5#2#6)}
\crefformat{figure}{figure~#2#1#3}
\crefformat{equation}{equation~(#2#1#3)}
\crefformat{table}{table~#2#1#3}

\AtBeginBibliography{\footnotesize{}}

\begin{document}

% List of changes
% \onecolumn
% \listofchanges
% \vfil
% \pagebreak
% \twocolumn

\title{Controlling the Wireless Power Transfer Mechanism of the Both-Sides Retrodirective System}

\author{Charleston Dale M. Ambatali,~\IEEEmembership{Member,~IEEE}, and Shinichi Nakasuka%, \\Bo Yang,~\IEEEmembership{Member,~IEEE}, and Naoki Shinohara,~\IEEEmembership{Senior Member,~IEEE}
        % <-this % stops a space
%\thanks{This paper was produced by the IEEE Publication Technology Group. They are in Piscataway, NJ.}% <-this % stops a space
\thanks{Manuscript received January 22, 2024; revised June 7, 2024. This work is supported in part by the Intelligent Space Systems Laboratory (ISSL), Department of Aeronautics and Astronautics, The University of Tokyo, and in part by Hitachi, Ltd., Japan. \textit{(Corresponding author: Charleston Dale M. Ambatali.)}

C. D. M. Ambatali was with the Department of Aeronautics and Astronautics, The University of Tokyo, Hongo, Bunkyo 113-0033, Japan. He is now with the Electrical and Electronics Engineering Institute, University of the Philippines, Diliman, Quezon City 1101, Philippines (e-mail: cmambatali@up.edu.ph).

S. Nakasuka is with the Department of Aeronautics and Astronautics, The University of Tokyo, Hongo, Bunkyo 113-0033, Japan (e-mail: nakasuka@space.t.u-tokyo.ac.jp).

%B. Yang and N. Shinohara are with the Research Institute for Sustainable Humanosphere, Kyoto University, Uji 611-0011, Japan (e-mail: yang\_bo@rish.kyoto-u.ac.jp; shino@rish.kyoto-u.ac.jp).
}}

\markboth{Journal of \LaTeX\ Class Files,~Vol.~14, No.~8, August~2021}%
{Shell \MakeLowercase{\textit{et al.}}: A Sample Article Using IEEEtran.cls for IEEE Journals}

\IEEEpubid{0000--0000/00\$00.00~\copyright~2024 IEEE}

\maketitle

\begin{abstract}
To achieve efficient long-range wireless power transfer (WPT), large antenna systems are necessary spanning lengths of tens to thousands of meters in one dimension.
This creates an array in the order of at least hundreds of thousands to billions of elements.
This makes the implementation of beamforming control a challenge.
Various works focus on iterative optimization or channel estimation to maintain high efficiency in a time-varying environment requiring complex processing capabilities.
A simpler alternative is the both-sides retrodirective antenna array (BS-RDAA) system where iterative optimization or channel estimation is not required. In a previous study, it was observed that this system achieves maximum WPT efficiency if the system is marginally stable. Thus, there is a need to regulate the system to maintain marginal stability regardless of the transmission channel conditions. In this paper, we present a plant model for the system and design a control system to drive it to marginal stability.
The result is a WPT implementation that is not dependent on complex processing capabilities present in other established high efficiency methods.
We also confirmed the ability of the proposed design to maintain maximum efficiency in a dynamic environment through the results of an electromagnetic simulator and a time domain simulator.

%In this paper, we present a discrete-time state space model of the aforementioned system and show that it can naturally achieve maximum theoretical WPT efficiency at steady-state.
\end{abstract}


\section{Introduction}
\label{sec:introduction}
    \IEEEPARstart{W}{ireless} power transfer (WPT) can be achieved in the order of hundreds of kilometers to tens of thousands of kilometers as long as the antenna size is large enough so that the system operates in the radiative near-field zone \cite{gowda2016}.
    It is in this region where electromagnetic waves are non-planar and impose an amplitude and phase distribution onto a planar receiver.
    A proper amplitude and phase distribution on the wave source will focus its beam on a planar receiver with minimal power lost to free space.
    Let $\lambda$ be the wavelength of operation which is related to the wave frequency, $f$, and the speed of light in free space, $c$, by the equation $c=\lambda f$.
    If $D$ is the maximum length of the antenna, then the maximum distance of the radiative near-field is computed by $2D^2/\lambda$ \cite{gowda2016,zhang2022nearfield}.
    If operated at 5.8 GHz, ground-based transmission systems with a distance of at most 10 km require antennas at least 16 m in maximum length.
    If a planar phased array with $\lambda/2$ spacing is used, there will be at least 380,000 elements.
    For comparison, massive multi-element antenna arrays used for communication systems require at least 100 elements \cite{lu2014massivemimo}, and the largest array so far has 4096 elements \cite{jordan2019antennaarray4096}.

    \IEEEpubidadjcol

    % \begin{figure}
    %     \centering
    %     \includegraphics[width=\linewidth]{images/fraunhoffer-distance.png}
    %     \caption{Minimum antenna length (m) vs. distance (km) for $5.8\;GHz$ where the blue line is the Fraunhoffer distance and the shaded region is the valid antenna sizes given the distance.}
    %     \label{fig:fraunhoffer-distance}
    % \end{figure}
    
    In the scale of space-based solar power (SBSP) from geostationary orbit ($\thicksim$36,000 $km$ distance) \cite{MANKINS2009146}, a minimum antenna length of 1 km is needed to be in the radiative near-field.
    From a previous work, we estimated that a 2.5 km$\times$2.5 km antenna array in orbit can focus a beam to a 5 km diameter on the ground \cite{ambatali2023comparison}.
    Using a $\lambda/2$ spacing between adjacent antenna elements arranged on a square plane, the number of antennas will be around 10 billion elements.
    % To achieve maximum WPTE, the generator array must have an amplitude and phase distribution determined through an eigendecomposition of the WPT channel \cite{yuan2023sparameters, choi2017receivedpowerestimation, ambatali2024characterizing}, specifically using the eigenvector associated with the maximum eigenvalue.
    Therefore, in these scales, beam control for maximum WPTE becomes a herculean task especially in a time-varying environment.

    Beam control methods for long distance WPT rely on systems to either track a target receiver or estimate the the S-parameters of the WPT channel, and compute for the optimal phase and amplitude on each antenna element on the generator array.
    Scaling them to a practical implementation from ground-based to space-based implementations is an active area of inquiry.
    The different methods that have been developed can be roughly categorized into three classes: 1) the conventional antenna array beam steering used in conjunction with position sensors \cite{yang2023autotracking, anselmi2023selfreplicating, hui2020directionalradiation}, 2) use of pilot signals to guide the generator antennas \cite{matsumuro_2017, koo2021retroreflectivearray}, or 3) feedback control using data sent back by the receiver to the generator \cite{hui2020directionalradiation, hajimiri2021dynamicbeam, choi2017receivedpowerestimation}.
    \Cref{tab:comparison-of-methods} summarizes a comparison of their performance and the added capability required on the receiver and generator arrays to implement them.
    On all the approaches, there is generally a trade-off observed between the WPTE performance and the convergence time needed to achieve their steady state.

    \begin{table*}
        \centering
        \caption{Different works on WPT control in terms of receiver capabilities, generator capabilities, and steady-state performance.}
        \label{tab:comparison-of-methods}
        \begin{tabular}{|p{0.195\linewidth}|c|p{0.15\linewidth}|p{0.165\linewidth}|p{0.165\linewidth}|p{0.145\linewidth}|}
            \hline
            \hfil\textbf{Method}\hfil & \hfil\textbf{Ref.}\hfil & \hfil\textbf{Type}\hfil & \hfil\textbf{Receiver Capabilities}\hfil & \hfil\textbf{Generator Capabilities}\hfil & \hfil\textbf{Performance}\hfil \\\hline
            Visual tracking with camera & \cite{yang2023autotracking} & Conventional beam steering & (none) & Visual tracking, precise phase control & Sub-optimal efficiency, fast convergence \\\hline
            Directional radiation using position information & \cite{hui2020directionalradiation} & Conventional beam steering & Position reporting & Position tracking, precise phase control & Sub-optimal efficiency, fast convergence \\\hline
            Focused pilot signal transmission & \cite{matsumuro_2017} & Pilot signal beam forming & Pilot signal transmission, phase control & Retrodirective antenna array & Sub-optimal efficiency, fast convergence \\\hline
            \textit{In-situ} antenna calibration & \cite{koo2021retroreflectivearray} & Pilot signal beam forming & Pilot signal transmission, phase control & Retrodirective phase control, digital processing & Sub-optimal efficiency, slow convergence \\\hline
            Iterative power optimization & \cite{hajimiri2021dynamicbeam} & Feedback control & Telemetry transmission & Processor capable of optimization, phase control & Sub-optimal efficiency, slow convergence \\\hline
            Channel estimation from receiver power & \cite{choi2017receivedpowerestimation} & Feedback control & Telemetry transmission & Signal processor for channel estimation, phase control, amplitude control & Maximum efficiency, slow convergence \\\hline
            Directional radiation by iterative superposition & \cite{hui2020directionalradiation} & Feedback control & Telemetry transmission & Simple signal processing, phase control, amplitude control & Maximum efficiency, slow convergence \\\hline
            Both-sides retrodirective antenna array & \cite{matsumuro2019basic} & Feedback control, Pilot signal beam forming & Retrodirective antennas, automatic gain control & Retrodirective antennas, automatic gain control & Maximum efficiency, fast convergence \\\hline
        \end{tabular}
    \end{table*}

    To achieve maximum WPTE, complete knowledge of the S-parameters of the wireless channel from power generator to receiver is required to compute the optimal amplitude and phase settings \cite{yuan2023sparameters, kato2023emimo}.
    Thus, it is imperative to continuously estimate the channel and compute the optimum transmission setting.
    This can be done through feedback \cite{choi2017receivedpowerestimation}.
    Alternatively, using iterative methods to tune each element over successive iterations based on an objective function \cite{hui2020directionalradiation, hajimiri2021dynamicbeam}, need for channel estimation is eliminated.
    However, they require more time to converge when more antenna elements are added which is not ideal for a dynamic wireless channel.
    To make long-range WPT scalable, there is a need for a method that converges or settles faster and does not require channel estimation.
    A candidate is the both-sides retrodirective antenna array (BS-RDAA) system in which the retrodirective capability is installed on both the generator side and the receiver side \cite{matsumuro2019basic}.
    We differentiate it from the conventional \textit{single-side} retrodirective antenna array (SS-RDAA) which installs retrodirective capability on the generator side only.
    The SS-RDAA can be classified as a pilot signal beamforming method of WPT.
    Although BS-RDAA also uses a pilot signal, a part of the received power is used to re-adjust this pilot signal which is akin to a feedback mechanism.

    An optimum pilot signal can bring out the maximum WPTE beam setting of the transmitter \cite{matsumuro2017beampilot}.
    Further analysis in a previous work by the authors reveals that in a marginally stable condition at steady state, the BS-RDAA naturally produces this optimum pilot signal leading to a maximum WPTE condition \cite{AMBATALI2024experimentalvalidation, ambatali2024characterizing}.
    It requires no prior knowledge of the channel and automatically adjusts its antenna excitations according to any changes.
    This makes the BS-RDAA concept less complex compared to existing methods while still achieving maximum WPTE provided that marginal stability is maintained.
    To achieve this condition and to be able to control the power output of the system, an appropriate control system must be used.
    
    In this paper, we present a simple plant model of the BS-RDAA system using the analysis from previous works as a basis.
    We also designed a controller using classical methods applied to the plant model.
    Specifically, we propose the use of a proportional integral (PI) controller using the power measured at the output as a feedback to control the loop gain of the BS-RDAA and maintain its marginal stability state.
    %we propose a proportional and integrator controller system designed through classical methods to maintain the BS-RDAA in the marginally stable state and stabilize its power output.
    Through the results of an electromagnetic simulation incorporated into a time domain simulation model, we verify that the proposed design can maintain the marginal stability state and regulate the power output.
    Thus, we present the completed BS-RDAA system, a WPT beaming technique that requires no estimation or knowledge of the channel, has the ability to adjust to changing conditions, and achieves maximum WPTE based on the real-time S-parameters of the channel.

    The rest of this paper is structured as follows.
    \Cref{sec:longrangewpt} discusses the calculation of wireless power transfer efficiency and compares the existing WPT beaming methods based on it.
    \Cref{sec:plant-model} discusses the dynamics of the BS-RDAA, the experiment done to verify the analysis, and the derivation of its plant model that is used for the design of the control system.
    Additionally, the proposed control system is also characterized in terms of its settling time, disturbance, and overshoot.
    \Cref{sec:simulation} presents the simulation setup used to verify the performance of the proposed controller design.
    The S-parameters of a reference WPT setup is calculated using an electromagnetic simulator which is then used to model the channel behavior in an ordinary differential equation (ODE) solver.
    Finally, \Cref{sec:conclusion} summarizes the paper and gives a recommendation on what further work must be done.

\section{Wireless Power Transfer Efficiency}
\label{sec:longrangewpt}
    \subsection{Mathematical Notations}
    Scalar quantities are written in italics, column vectors in lower case bold, and matrices in uppercase bold.
    For example, $a$ and $A$ are scalars, $\mathbf{a}$ is a column vector, and $\mathbf{A}$ is a matrix.
    The notation $\{a_i\}$ indicates a column vector for a positive integer set defined by $i$ and $\{A_{ij}\}$ is a matrix with a column index $i$ and a row index $j$, both positive integers.
    Since these are complex quantities, we then define $|a|$ and $\angle a$ as the magnitude and phase of $a$, respectively.
    
    The transpose and Hermitian transpose operations of $\mathbf{a}$ are defined by $\mathbf{a}^T$ and $\mathbf{a}^H$, respectively, and these operations transform a column vector into a row vector and vice versa.
    The per-element conjugation operation is denoted as $\mathbf{\Bar{a}}$.
    The three operations outlined above can also be applied to matrices.
    The conjugation operation can also be done on scalars.

    We denote the superscript $(k)$ as the discrete-time index.
    The value of some quantity $a$, $\mathbf{a}$, or $\mathbf{A}$ at some time $(k)$ is denoted by $a^{(k)}$, $\mathbf{a}^{(k)}$, and $\mathbf{A}^{(k)}$, respectively.
    We differentiate this with the exponentiation notation $a^k$ where $a$ is a base and $k$ is an exponent.
    We denote $\mathbb{R}$ and $\mathbb{C}$ as the set of all real numbers and complex numbers, respectively.
    Finally, we define $u(t)$ as the unit Heaviside step function.
    
    \subsection{S-parameter model of WPTE}

    % S-parameters are used to characterize how a voltage wave is scattered in a microwave circuit with an arbitrary number of ports.
    % It is widely used to characterize high frequency circuits and has been used to characterize waveguides -- devices that uses electromagnetic fields to propagate information \cite{pozar2011microwavech04}.
    % The wireless channel can also be imagined as a waveguide without boundaries, thus, an antenna system communicating via the wireless channel can be expressed by S-parameters.

\begin{figure}
        \centering
        \begin{circuitikz}[american voltages,scale=0.9]
    \draw[dashed] 
        (-2,-1.25) rectangle (2,3.25)
        (-0.75,0) rectangle (3.25,4.5)
        (-2,-1.25) -- (-0.75,0)
        (2,3.25) -- (3.25,4.5)
        (-2,3.25) -- (-0.75,4.5)
        (2,-1.25) -- (3.25,0)
    ;
    
    \draw 
        (-3,2.5) to[short,o-] (-1.5,2.5) node[bareantenna,rotate=-90]{}
        (-3,0.5) to[short,o-] (-1.5,0.5) node[ground]{}
        (-3,2.5) to[open,v=$\mathbf{v_1}$] (-3,0.5)
        (-2.75,2.75) to[short,o-] (-1.25,2.75) node[bareantenna,rotate=-90]{}
        (-2.75,0.75) to[short,o-] (-1.25,0.75) node[ground]{}
        (-1.75,3.75) to[short,o-] (-0.25,3.75) node[bareantenna,rotate=-90]{}
        (-1.75,1.75) to[short,o-] (-0.25,1.75) node[ground]{}
        
        (3,2.5) to[short,o-] (1.5,2.5) node[bareantenna,rotate=90]{}
        (3,0.5) to[short,o-] (1.5,0.5) node[ground]{}
        (4.25,3.75) to[open,v=$\mathbf{v_2}$] (4.25,1.75)
        (3.25,2.75) to[short,o-] (1.75,2.75) node[bareantenna,rotate=90]{}
        (3.25,0.75) to[short,o-] (1.75,0.75) node[ground]{}
        (4.25,3.75) to[short,o-] (2.75,3.75) node[bareantenna,rotate=90]{}
        (4.25,1.75) to[short,o-] (2.75,1.75) node[ground]{}

        (-4,1.125) node[rotate=90]{\textbf{Rectenna Side}}
        (5,1.125) node[rotate=-90]{\textbf{Generator Side}}
    ;
    \foreach \x in {0,...,2}
        \draw
            (-2.5+0.25*\x,1+0.25*\x) node[circ]{}
            (3.5+0.25*\x,1+0.25*\x) node[circ]{}
            (-2.5+0.25*\x,3+0.25*\x) node[circ]{}
            (3.5+0.25*\x,3+0.25*\x) node[circ]{}
        ;
\end{circuitikz}
        \caption{The WPT system represented as a network of antennas which can be characterized by network parameters such as Z-, Y-, or S-parameters.}
        \label{fig:black-box-wpt-system}
    \end{figure}
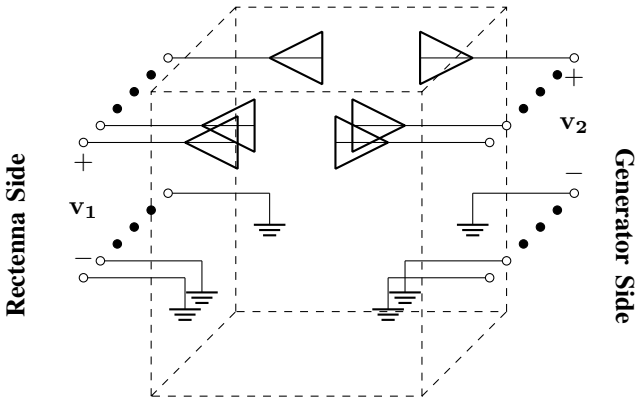

    For a system of $N$ power generating antennas and $M$ receiver antennas shown in \cref{fig:black-box-wpt-system}, let $\mathbf{v_1}\in\mathbb{C}^{M\times1}$ be the voltage at the ports on the receiver array and $\mathbf{v_2}\in\mathbb{C}^{N\times1}$ as those on the generator array \cite{yuan2023sparameters}.
    Conveniently, the total power at the receiver and the transmitter can be expressed by $\mathbf{v_1}^H\mathbf{v_1}/Z_0$ and $\mathbf{v_2}^H\mathbf{v_2}/Z_0$, respectively, where $Z_0$ is the matched impedance assumed to be the same for all ports.
    The relationship between the voltages can be defined by the S-parameters shown in \cref{eq:black-box-wpt} where $\mathbf{v_1=v_{1f}+v_{1b}}$ and $\mathbf{v_2=v_{2f}+v_{2b}}$.
    Subscript $\mathbf{f}$ indicates a forward wave, or a wave going in, and $\mathbf{b}$ means backward wave, or a wave going out of the system and into the respective ports.
    \begin{equation}
        \begin{pmatrix}
            \mathbf{v_{1b}} \\ \mathbf{v_{2b}}
        \end{pmatrix}=\begin{pmatrix}
            \mathbf{S_{11}} & \mathbf{S_{12}} \\
            \mathbf{S_{21}} & \mathbf{S_{22}} \\
        \end{pmatrix}\begin{pmatrix}
            \mathbf{v_{1f}} \\ \mathbf{v_{2f}}
        \end{pmatrix}
        \label{eq:black-box-wpt}
    \end{equation}

    In \cref{eq:black-box-wpt}, the diagonal submatrices, $\mathbf{S_{11}}$ and $\mathbf{S_{22}}$, describe the wave reflected back to their respective set of ports.
    Ideally, all their elements are zero.
    On the other hand, the non-diagonal submatrices indicate the waves that is transmitted from one set of ports to another.
    For example, $\mathbf{S_{12}}$ represents the power sent from $\mathbf{v_2}$ to $\mathbf{v_1}$ and this is where we want power to flow.
    The wireless channel is assumed to be reciprocal implying that $\mathbf{S_{12}}=\mathbf{S_{21}}^T$.

    If there is an input at the power generators, the formula for the WPTE can be expressed as the ratio of the power at the receiver from $\mathbf{v_{1b}=S_{21}}^T\mathbf{v_{2f}}$ to the input power at the generator from $\mathbf{v_{2f}}$ and this is expressed in \cref{eq:rayleigh-quotient-formulation}.
    This is called the Rayleigh coefficient which is widely used in different works when maximizing the WPTE of antenna arrays \cite{hajimiri2021dynamicbeam, oliveri2013maxbeam}.
    The maxima of this is the maximum eigenvalue of $\mathbf{\Bar{S}_{21}}\mathbf{S_{21}}^T$.
    Let $\xi$ be the eigenvalues of the matrix expression.
    Since $\mathbf{\Bar{S}_{21}}\mathbf{S_{21}}^T$ is Hermitian, then $\xi\in\mathbb{R}$ for all eigenvalues.
    Then the maximum theoretical efficiency is governed by the largest possible $\xi$ as expressed in \cref{eq:max-efficiency}.
    
    \begin{equation}
        \eta = \frac{\mathbf{v_{2f}}^H\left(\mathbf{\Bar{S}_{21}}\mathbf{S_{21}}^T\right)\mathbf{v_{2f}}}{\mathbf{v_{2f}}^H\mathbf{v_{2f}}}
        \label{eq:rayleigh-quotient-formulation}
    \end{equation}
    
    \begin{equation}
        \eta_{max}=\max\left(\mathrm{eig}\left\{\mathbf{\Bar{S}_{21}}\mathbf{S_{21}}^T\right\}\right)=\xi_{max}
        \label{eq:max-efficiency}
    \end{equation}

    An arbitrary input to the generator array can be expressed as a weighted sum of the set of the normalized eigenvectors of $\mathbf{\Bar{S}_{21}}\mathbf{S_{21}}^T$. This is shown in \cref{eq:eigen-expression-s12} where $v_{2i}\in\mathbb{C}$ and $\mathbf{a}_i$ is an eigenvector for all $i\in\{1,2,...,M\}$.
    This assumes that $\mathbf{\Bar{S}_{21}}\mathbf{S_{21}}^T$ is a full rank matrix.
    \begin{equation}
        \mathbf{v_{2f}}=\sum_{i=1}^M v_{2i}\mathbf{a}_i
        \label{eq:eigen-expression-s12}
    \end{equation}

    Now, the efficiency of any arbitrary input can be expressed in terms of the weights, $v_{2i}$, and eigenvalues, $\xi_i$, as shown in \cref{eq:efficiency-weights-eigenvalue}.
    This also shows that each $\xi_i$ is the WPTE of the corresponding beam mode, $\mathbf{a}_i$, and the efficiency is a weighted average of the individual efficiencies of each beam mode.
    \begin{equation}
        \eta=\frac{\sum\limits_{i=1}^M\xi_i|v_{2i}|^2}{\sum\limits_{i=1}^M|v_{2i}|^2}
        \label{eq:efficiency-weights-eigenvalue}
    \end{equation}
    
    A scalar multiple of the eigenvector associated with $\xi_{max}$, denoted by $\mathbf{a}_{max}$, is the optimum setting for the input.
    In the case that multiple eigenvectors are associated with $\xi_{max}$, then the optimum input can be a linear combination of the eigenvectors.
    The power can be controlled by the magnitude of the weights.
    To achieve high efficiency WPT, a system must be designed such that the maximum WPTE condition, $\mathbf{v_{2f}}=c\mathbf{a}_{max}$, is maintained as the channel changes over time.

\section{BS-RDAA Plant Model}
\label{sec:plant-model}
    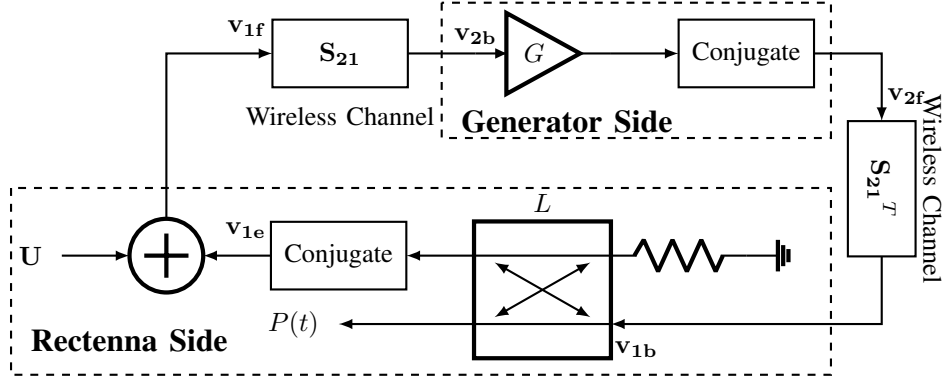
\begin{figure*}
        \centering
        \begin{circuitikz}[american voltages, american currents, thick, scale=0.9]
    \coordinate (s21) at (0,0);
    \draw [-] ($(s21)+(2,0)$) -- ++(1,0) coordinate(amp1);
    
    \draw
        ($(s21)-(0,0.5)$) rectangle ($(s21)+(2,0.5)$)
        ($(s21)+(1,0)$) node[]{$\mathbf{S_{21}}$}
        ($(s21)+(-0.375,0.375)$) node[]{$\mathbf{v_{1f}}$}
        ($(s21)+(1,-0.5)$) node[label={below:Wireless Channel}]{}

        (amp1) to[amp,>,t=$G$] ++(2,0) coordinate(amp2)
        (amp1) node[above]{$\mathbf{v_{2b}}$}
    ;

    \draw [-latex] (amp2) -- ++(1,0) coordinate(conj1);
    \draw [-latex] ($(conj1)+(2,0)$) -- ++(1,0) -- ++(0,-1) coordinate(s12);
    \draw [-latex] ($(s12)-(0,2)$) -- ++(0,-1) -- ++(-4,0) coordinate(cplr);

    \draw
        ($(conj1)-(0,0.5)$) rectangle ($(conj1)+(2,0.5)$)
        ($(conj1)+(1,0)$) node[]{Conjugate}

        ($(s12)+(0.5,0)$) rectangle ($(s12)+(-0.5,-2)$)
        ($(s12)-(0,1)$) node[rotate=-90]{$\mathbf{S_{21}}^T$}
        ($(s12)+(0.375,0.375)$) node[]{$\mathbf{v_{2f}}$}
        ($(s12)-(-0.75,1)$) node[rotate=-90]{Wireless Channel}

        (cplr) node[coupler,anchor=right down](cplrnode){$L$}
        (cplrnode.right up) to[R] ++(2,0) node[ground,rotate=90]{}
        ($(cplr)-(-0.375,0.375)$) node[]{$\mathbf{v_{1b}}$}
    ;

    \draw [-latex] (cplrnode.left down) -- ++(-2,0) coordinate(powerout);
    \draw [-latex] (cplrnode.left up) -- ++(-1,0) coordinate(conj2);
    \draw [-latex] ($(conj2)-(2,0)$) -- ++(-1,0) coordinate(adder1);
    
    \draw (adder1) node[adder,anchor=e](adder1node){};
    \draw [latex-] (adder1node.w) -- ++(-1,0) node[label={left:$\mathbf{U}$}](input){};
    \draw [-latex] (adder1node.n) -- (adder1node.n |- s21) -- (s21);
    \draw
        (powerout) node[label={left:$P(t)$}]{}
        
        ($(conj2)+(0,0.5)$) rectangle ($(conj2)-(2,0.5)$)
        ($(conj2)-(1,0)$) node[]{Conjugate}
        ($(conj2)-(2.375,-0.375)$) node[]{$\mathbf{v_{1e}}$}

        ($(input)+(-0.75,-1.25)$) node[label={right:\large\textbf{Rectenna Side}}]{}
        ($(amp1)+(-0.5,-1)$) node[label={right:\large\textbf{Generator Side}}]{}
    ;

    \draw[dashed] ($(input)+(-0.75,1)$) rectangle ($(cplr) + (3.25,-0.75)$);
    \draw[dashed] ($(amp1)+(-0.5,0.75)$) rectangle ($(s12) + (-0.75,-0.25)$);
\end{circuitikz}
        \caption{Block diagram of a both-sides retrodirective system with an input $\mathbf{U}$. In the generator side, the power is boosted by gain $G$ and conjugated. In the receiver side, most of the power is collected in the coupler's through path and part of it is fed back to the system with loss $L$. The output power is $P(t)$.}
        \label{fig:bsrdaa-block-diagram}
        \vspace{-0.4cm}
    \end{figure*}

    \subsection{Dynamics of BS-RDAA}
    
    The block diagram of the BS-RDAA concept is shown in \cref{fig:bsrdaa-block-diagram}, assuming negligible $\mathbf{S_{11}}$ and $\mathbf{S_{22}}$.
    From the receiver side, a pilot signal vector, $\mathbf{v_{1f}}$, is sent passing through the channel.
    This is received as $\mathbf{v_{2b}=S_{21}v_{1f}}$ by the generator antennas.
    Power is injected through the gain, $G$, and is conjugated as per retrodirective operation \cite{miyamoto2002retrodirective}.
    In this system, $G$ should be equal in magnitude in all elements.
    The output of the generator, $\mathbf{v_{2f}}$, comes back to the receiver through the same channel received as $\mathbf{v_{1b}=S_{21}}^T\mathbf{v_{2f}}$.
    Finally, the receiver array passes $\mathbf{v_{1b}}$ through a coupler at each element where majority of the power, $P(t)$, passed to a power grid.
    A small part of it, attenuated by loss $L$, is conjugated and fed back to the system as a new pilot signal and the cycle continues.
    The input $\mathbf{U}$ is added for generalization of the block diagram.
    It can represent an impulse signal to jump start the system or white noise present in the electronics of the system.

    The discrete-time characteristic equation of \cref{fig:bsrdaa-block-diagram} can be expressed as \cref{eq:characteristic-equation}, where one sampling period is assumed to be the sum of the round-trip time of the signal from the receiver side to the generator side and the processing delay of the other system components, namely, the conjugators, couplers, and amplifiers.
    
    \begin{equation}
        \begin{split}
            \mathbf{v^{\mathnormal{(k)}}_{1f}}=\Bar{L}G\mathbf{S_{21}}^H\mathbf{S_{21}v^{\mathnormal{(k-1)}}_{1f}}+\mathbf{U}^{(k)}, \\
            \mathbf{v^{\mathnormal{(k)}}_{2f}}=L\Bar{G}\mathbf{\Bar{S}_{21}}\mathbf{S_{21}}^T\mathbf{v^{\mathnormal{(k-1)}}_{2f}}+\Bar{G}\mathbf{\Bar{S}_{21}}\mathbf{\Bar{U}}^{(k)}.
        \end{split}
        \label{eq:characteristic-equation}
    \end{equation}
    
    % Before proceeding, we first express $\mathbf{S_{21}}$ by its singular value decomposition in \cref{eq:singular-value-expression}.
    % By definition, $\mathbf{A}\in\mathbb{C}^{N\times N}$ and $\mathbf{B}\in\mathbb{C}^{M\times M}$ are unitary matrices, and $\mathbf{\Lambda}\in\mathbb{C}^{N\times M}$ is a rectangular diagonal matrix consisting of the singular values of $\mathbf{S_{21}}$.
    % \begin{equation}
    %     \mathbf{S_{21}}=\mathbf{A\Lambda B}^H
    %     \label{eq:singular-value-expression}
    % \end{equation}
    
    % The column vectors of $\mathbf{\Bar{A}}$ and $\mathbf{B}$ are the normal eigenvectors of $\mathbf{\Bar{S}_{21}S_{21}}^T$ and $\mathbf{S_{21}}^H\mathbf{S_{21}}$, respectively.
    % Then \cref{eq:eigenvalue-decomposition} reveals that the maximum number of eigenvalues is defined by $\min(M,N)$.
    % It also shows that the eigenvalues for both cases are the same and that they are real.
    % Without loss of generality, we assume $M<N$.
    % \begin{equation}
    %     \begin{split}
    %         \mathbf{S_{21}}^H\mathbf{S_{21}}=\mathbf{B\Lambda}^H\mathbf{\Lambda B}^H \\
    %         \mathbf{\Bar{S}_{21}S_{21}}^T=\mathbf{\Bar{A}\Bar{\Lambda}\Lambda}^T\mathbf{A}^T
    %     \end{split}
    %     \label{eq:eigenvalue-decomposition}
    % \end{equation}
    
    The zero-input response, i.e. $\mathbf{U=0}$, for an arbitrary initial condition is given by \cref{eq:zero-input-response}.
    Here, $\mathbf{b}_i$ is an eigenvector of $\mathbf{S_{21}}^H\mathbf{S_{21}}$ and $v_{1i}$ are weights for $\mathbf{v_{1f}}$, while $v_{2i}$ and $\mathbf{a}_i$ holds the same definition as that of \cref{eq:eigen-expression-s12}.
    The set of eigenvectors, $\mathbf{b}_i$, is parallel to the physical aspect of $\mathbf{a}_i$, but it represents the beam modes from the receiver to the generator.
    \begin{equation}
        \begin{split}
            \mathbf{v^{\mathnormal{(k)}}_{1f}}=(\Bar{L}G)^k\sum_{i=1}^M\xi_i^kv_{1i}^{(0)}\mathbf{b}_i, \\
            \mathbf{v^{\mathnormal{(k)}}_{2f}}=(L\Bar{G})^k\sum_{i=1}^M\xi_i^kv_{2i}^{(0)}\mathbf{a}_i.
        \end{split}
        \label{eq:zero-input-response}
    \end{equation}
    
    Marginal stability is achieved when the steady state value is finite and non-zero as $k$ approaches infinity.
    This is satisfied when the value $|LG|$ is the reciprocal of $\xi_{max}$ as expressed in \cref{eq:marginal-stability-linear}.
    $|LG|>\xi_{max}^{-1}$ will render the system unstable while $|LG|<\xi_{max}^{-1}$ makes it stable.
    The resulting steady state expressions in marginal stability are given by \cref{eq:zero-input-steady-state}.
    \begin{equation}
        \xi_{max}=\eta_{max}=\frac{1}{|LG|}
        \label{eq:marginal-stability-linear}
    \end{equation}
    \begin{equation}
        \begin{split}
            \lim_{k\xrightarrow{}+\infty}\mathbf{v^{\mathnormal{(k)}}_{1f}}=(\Bar{L}G\xi_{max})^kv_{1,max}^{(0)}\mathbf{b}_{max}, \\
            \lim_{k\xrightarrow{}+\infty}\mathbf{v^{\mathnormal{(k)}}_{2f}}=(L\Bar{G}\xi_{max})^kv_{2,max}^{(0)}\mathbf{a}_{max}.
        \end{split}
        \label{eq:zero-input-steady-state}
    \end{equation}
    
    Since steady state value of $\mathbf{v_{2f}}$ is a scalar multiple of $\mathbf{a}_{max}$, computing the efficiency according to \cref{eq:rayleigh-quotient-formulation} results to $\eta=\xi_{max}$.
    Therefore, achieving marginal stability on a both-sides retrodirective system achieves maximum theoretical efficiency for WPT.

    \subsection{Experimental verification of analysis}

\begin{figure}
        \centering
        \includegraphics[width=\linewidth]{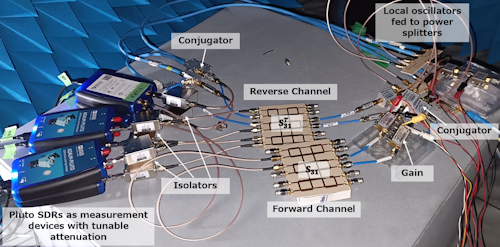}
        \caption{Experiment setup to verify the theoretical analysis of the behavior of the BS-RDAA system.}
        \label{fig:bs-rdaa-experiment-image}

        \includegraphics[width=\linewidth]{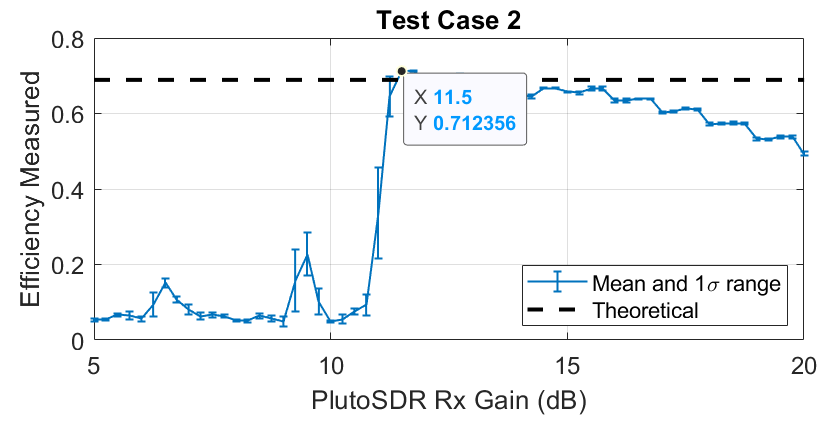}
        \caption{Efficiency measured versus gain setting for one test case.}
        \label{fig:test-case2}
        \vspace{-0.5cm}
    \end{figure}

    In previous works \cite{AMBATALI2024experimentalvalidation}, we verified the behavior of BS-RDAA derived in the previous subsection through an experiment setup shown in \cref{fig:bs-rdaa-experiment-image} which is an implementation of the block diagram in \cref{fig:bsrdaa-block-diagram}.
    A 12-port circuit board is used as a representation of a static wireless channel with 6 ports representing the generator side and the other 6 ports as the receiver side.
    Its $12\times12$ S-parameter matrix at 2.4 GHz is measured to compute for the theoretical efficiency at different test cases.
    A single test case is using a combination of two ports from the transmitter side and two ports from the receiver side.
    All unused ports are terminated by a 50 $\Omega$ load to model the power lost to free space.
    In each test case, the loop gain $|LG|$ is swept from a high value to a low value to observe the unstable, marginally stable, and stable cases predicted by theory.
    Finally, thirty (30) test cases were checked. The loop gain at marginal stability, theoretical efficiency, and measured efficiency were all noted.

\begin{figure}
        \includegraphics[width=\linewidth]{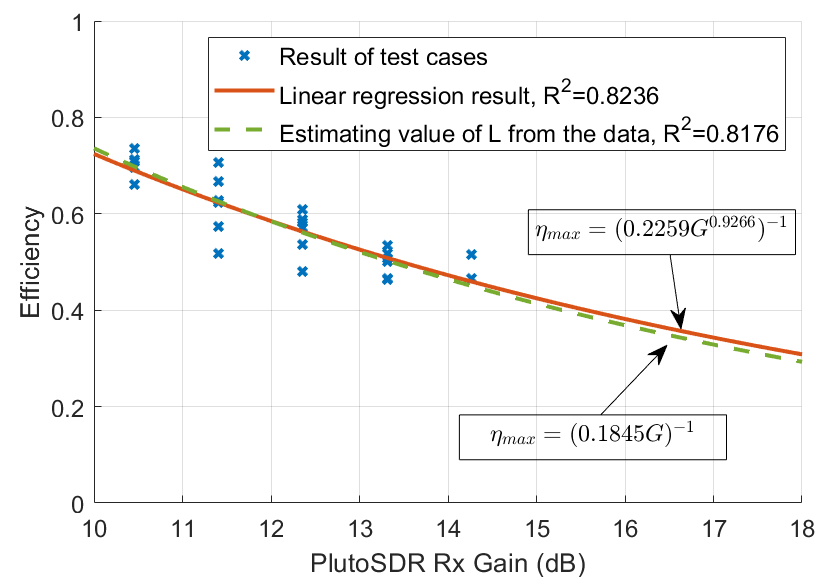}
        \caption{Linear regression performed on the thirty (30) test cases to verify the marginal stability condition. }
        \label{fig:marginal-stability}
    \end{figure}

    From \cref{eq:zero-input-response}, if the BS-RDAA system is stable, the output will go down to zero and the efficiency that will be measured is due to noise.
    If the system is unstable, the output theoretically approaches infinity.
    In a practical setting, the receiver electronics will be driven into saturation.
    This behavior are observed in \cref{fig:test-case2} where when the gain is low, the measured efficiency is dictated by noise while when the gain is high, the measured efficiency decreases due to more receivers becoming saturated.
    In between these two regions is the point of marginal stability in which the efficiency is at its peak.
    This behavior is also confirmed by \cref{fig:marginal-stability} where \cref{eq:marginal-stability-linear} is verified.
    From the graph, a linear regression curve (red) is computed in the logarithmic domain, and one curve whose $L$ is computed from the data (green).
    Both curves have an $R^2$ greater than 0.8 which indicates a strong confirmation of the predicted marginal stability condition in \cref{eq:marginal-stability-linear}.
    Thus, the predicted behavior of the BS-RDAA system was verified.

%\section{Power Output Control}
%\label{sec:control-design}    
    \subsection{Plant model characteristics and limitations}
    The power output at the receiver side is computed with $\mathbf{v_{1b}}^H\mathbf{v_{1b}}/Z_0$ and $\mathbf{v_{1b}=S_{21}}^T\mathbf{v_{2f}}$.
    Using \cref{eq:zero-input-steady-state} for $\mathbf{v_{2f}}$, the power over time is simplified to \cref{eq:power-continuous-time} where $P_0$ is the initial power determined by the quantities $v_{2,max}^{(0)}$, $\xi_{max}$, $L$, $G$, and $Z_0$.
    
    \begin{equation}
        \label{eq:power-continuous-time}
        P^{(k)}=P_0\left|\xi_{max}\Bar{L}G\right|^{2k}
    \end{equation}

    To convert to continuous-time and apply the methods of Classical Control Theory, we set $k=t/t_p$ where $t_p$ is the loop propagation time.
    This term factors in the two-way propagation delay between receiver and generator and circuit processing delay in the system.
    In this domain, the BS-RDAA can remain purely analog in nature reducing the complexity of implementation.
    
    To linearize the system, we use the logarithm of \cref{eq:power-continuous-time}.
    Let $y=10\log{P}$, $y_0=10\log{(P_0)}$, $r=20\log\left|L\right|$, and $g=20\log\left|\xi_{max}G\right|$.
    The units of $y(t)$ and $y_0$ are decibel-watts (dBW) while $r$ and $g$ are in decibels (dB).
    The result is a linear equation expressed in \cref{eq:power-logarithmic-time}.
    %Initially, if there is no power, $y_0$ will be defined by additive white Gaussian noise power, i.e., it will never approach $-\infty$ in a practical setting.
    \begin{equation}
        \label{eq:power-logarithmic-time}
        y(t)=r\frac{t}{t_p}+g\frac{t}{t_p}+y_0
    \end{equation}

    The above equation is reminiscent of the position-velocity relationship where $y(t)$, the power, represents position, and $r$ or $g$, the loss and gain terms, respectively, collectively represent the velocity.
    From this, the plant model can be represented by an integrator with coefficient equal to $1/t_p$.
    Variables $r$ or $g$ can be controlled by using a voltage variable attenuator like the Mini Circuits RVA-3000R+ and similar devices.
    Without loss of generality, we can assume $r$ as the control signal and $g$ as the disturbance signal.

    The measurement of $y$ can be used as feedback but it must be made in the logarithmic domain in which radio frequency power meters are capable of performing.
    If a more inexpensive solution is desired, a logarithmic amplifier can be used as a sensor measuring the rectified output instead, but the RF-to-DC conversion loss must be taken into account.
    Given that the gain can be controlled and the power can be measured with existing devices, the originally multiple-input multiple-output control problem becomes a single-input single-output system.

    %Since the quantities are in the logarithm space, then $y_0$ is negative at the start.
    In theory, if there is no power initially (i.e. $P_0=0$), then $y_0\xrightarrow{}-\infty$ in which no meaningful power will be observed at the output as well.
    This is expected as the system relies on marginal stability to function.
    In practical systems, however, electronic noise is present, thus $y_0$ has a minimum value defined by the noise power.
    Consequently, the power supply can rise from this initial condition.

    A limitation of the proposed control system model model is defined by the dynamic range of $r$ in a real-world implementation.
    Particularly, it limits the channel conditions, defined by $\xi_{max}$, that the system can support as a result of \cref{eq:marginal-stability-linear}.
    For example, the Mini Circuits RVA-3000R+ voltage variable attenuator can only operate from $-35\;dB$ to $-3\;dB$ gain, corresponding to $1.41\%<\xi_{max}<56.23\%$ for $G=40\;dB$.
    Operating near the minimum and maximum capability of the controller should be avoided to maintain the desired settling time of the system.
        
    \subsection{Controller design}
    
\begin{figure} [!b]
        \centering
        \begin{circuitikz}[american voltages, american currents, thick]
    \coordinate (origin) at (0,0);

    \draw (origin) node[adder,scale=0.5](adder1){};
    \draw[latex-] (adder1.west) -- ++(-0.5,0) node[label={left:$R(s)$}]{};
    \draw[-latex] (adder1.east) -- ++(0.5,0) coordinate(icontrollerwest);
    \draw (icontrollerwest) to[open] ++(0.5,0) coordinate (icontroller);

    % Integral Controller
    \draw ($(icontroller)-(0.5,0.5)$) rectangle ($(icontroller)+(0.5,0.5)$);
    \draw (icontroller) node[]{\Large $\frac{K_I}{s}$};
    %\draw ($(icontroller)-(0,0.5)$) node[label={below: I Control}]{};
    \draw ($(icontroller)+(1.5,0)$) node[adder,scale=0.5](adder2){};
    \draw [-latex] ($(icontroller)+(0.5,0)$) -- (adder2.west);
    \draw[-latex] (adder2.east) -- ++(0.5,0) coordinate(plantwest);
    \draw (plantwest) to[open] ++(0.5,0) coordinate (plant);

    % Plant
    \draw ($(plant)-(0.5,0.5)$) rectangle ($(plant)+(0.5,0.5)$);
    \draw (plant) node[]{\Large $\frac{1}{st_p}$};
    \draw [dashed,thin] ($(adder2.north)+(-0.625,1)$) rectangle ($(plant)+(0.75,-1.5)$);
    \coordinate (disturbance) at ($(plantwest)+(0,0.875)$);
    \draw (disturbance) node[label={right:$G(s)$}]{};
    \draw [latex-] (adder2.north) -- (adder2.north |- disturbance) -- (disturbance);
    \draw ($(plant)+(0.5,-1)$) node[label={left:\textbf{PLANT}}]{};

    % Bias
    \coordinate (bias) at ($(icontroller)-(0,1)$);
    \draw (bias) node[label={left:$b$}]{};
    \draw[latex-] (adder2) -- ++(-1,-1) -- (bias);

    % Output
    \draw ($(plant)+(0.5,0)$) to[short,-*] ++(0.5,0) coordinate (outputnode);
    \draw[-latex] (outputnode) -- ++(0.5,0);
    \draw ($(outputnode)+(0.5,0)$) node[label={right:$Y(s)$}]{};

    % Feedback Gain
    \draw (outputnode) -- ++(0,-2.5) coordinate (feedback);
    \draw [-latex] (feedback) to[amp,>,t={$K_F$}] (adder2.south |- feedback) -- (adder2.south);
    \draw (outputnode |- feedback) to[short,*-] ++(0,-1) coordinate (feedback2);
    \draw [-latex] (feedback2) -- (adder1.south |- feedback2) -- (adder1.south);
    \draw ($(adder2.south)-(-0.25,0.25)$) node[]{$-$};
    \draw ($(adder1.south)-(-0.25,0.25)$) node[]{$-$};

    \draw [dashed,thin] ($(adder1.west)+(-0.25,1.25)$) -- ($(icontroller)+(0.625,1.25)$) -- ($(icontroller)+(0.625,-1.75)$) -- ($(feedback)+(0.25,0.75)$) -- ($(feedback2)+(0.25,-0.25)$) -- ($(feedback2)+(-5.5,-0.25)$) -- ($(adder1.west)+(-0.25,1.25)$);
    \draw (icontroller |- feedback) to[open] ++(0.125,0) node[]{\textbf{CONTROLLER}};
\end{circuitikz}
        \caption{Proposed control system block diagram with control parameters: feedback gain $K_F$, integral controller coefficient $K_I$, and bias $b$.}
        \label{fig:control-block-diagram}
        %\vspace{-0.4cm}
    \end{figure}
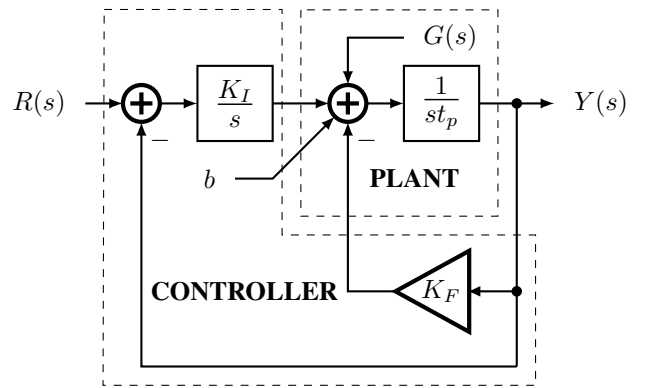

    \Cref{fig:control-block-diagram} shows the proposed control system design with three parameters: 1) the plant feedback gain $K_F$, 2) the integral controller coefficient $K_I$, and 3) the bias $b$.
    The parameter $K_F$ is used to control the speed in which $y(t)$ rises or falls while $K_I$ is used for the settling time of the step response of the system.
    The bias $b$ serves to counter the effect of $G(s)$ at the system start-up preventing overshoot.

    The input to output transfer function can be calculated as shown in \cref{eq:transfer-function-input} while the disturbance to output transfer function can be expressed by \cref{eq:transfer-function-disturbance}.
    Evaluating the final value of both expressions assuming step inputs show that the output signal is not influenced by the disturbance while also being equal to the reference input.
    
    \begin{equation}
        \label{eq:transfer-function-input}
        H_R(s)=\frac{Y(s)}{R(s)}=\frac{K_I}{t_ps^2+K_Fs+K_I}
    \end{equation}
    \begin{equation}
        \label{eq:transfer-function-disturbance}
        H_D(s)=\frac{Y(s)}{D(s)}=\frac{s}{t_ps^2+K_Fs+K_I}
    \end{equation}

    A transient quantity that must be considered is the non-zero initial condition $y_0$.
    It is expected to be comparable to the noise power which can be as low as $-140\;dBW$ at a temperature of $300\;K$ and $1\;MHz$ bandwidth.
    \Cref{eq:transfer-function-zero-input} defines the Laplace transform of the zero-input response for some constant initial condition $y_0$.

    \begin{equation}
        \label{eq:transfer-function-zero-input}
        Y_{Z}(s)=\frac{t_ps}{t_ps^2+K_Fs+K_I}y_0
    \end{equation}
    
    It is important to note that the output power is measured in decibel-watts which means that the power output grows exponentially by a factor of $10$ for every increase of $10$ of $y(t)$.
    Care must be given to any possible overshooting or fluctuating behavior which can stem from the disturbance or the non-zero initial condition in \cref{eq:transfer-function-disturbance,eq:transfer-function-zero-input}, respectively.
    %The design of the controller must be to calculate $K_F$, $K_I$, and $b$ to satisfy the settling time and overshoot requirements.
    
    \subsection{Step response behavior and stability}
    \Cref{eq:transfer-function-input} shows that the system is a second order system with two poles expressed in \cref{eq:poles-expression}.
    The system can be designed to be underdamped, critically damped, or overdamped.
    The underdamped case is not considered since overshooting is present.
    \begin{equation}
        \label{eq:poles-expression}
        p=\frac{-K_F\pm\sqrt{K_F^2-4K_It_p}}{2t_p}
    \end{equation}

    By setting $K_F>0$ and $K_I>0$, the real part of the poles will always be negative so a stable steady state output is guaranteed.
    %In the overdamped case, when $K_F^2>4K_It_p$, an upper bound approximation of the slower pole is $p\approx -K_I/K_F$.
    The critically damped case has a single pole at $p=-K_F/2t_p$ when $K_F=4K_It_p$ which predicts a the fastest settling time expressed in \cref{eq:settling-time}.
    The dependence of $t_s$ with $t_p$ is expected due to the physical limitations of the dynamic behavior of the BS-RDAA system.
    \begin{equation}
        \label{eq:settling-time}
        t_s=\frac{10t_p}{K_F}
    \end{equation}

    % Thus, we set the upper limit of $K_F$ to be unity.
    % From \cref{fig:settling-time-distance}, we expect the settling time of SBSP to be in the order of milliseconds in LEO and seconds in GEO.

    % \begin{figure}
    %     \centering
    %     \includegraphics[width=\linewidth]{images/settling-time-distance.png}
    %     \caption{Logarithmic plot of the best possible settling time as a function of distance between the receiver and generator antennas}
    %     \label{fig:settling-time-distance}
    % \end{figure}

    % Since $t_p$ can change over time, depending on the distance between the receiver and generator arrays, then trying to achieve $K_F^2=4K_It_p$ is impractical.
    % A more predictable design arises in the overdamped case in which the slower pole is approximated by only $K_F$ and $K_I$ which are control design parameters.
    % It is just important to note that the critically damped case is the fastest possible settling time and trying to achieve a design below that will result in overshooting.
    
    \subsection{Disturbance analysis}
    The term $G(s) $ is the Laplace transform of $g=20\log{\left|G\xi_{max}\right|}$. It is composed of the gain $G$ and the channel efficiency $\xi_{max}$. $G$ is the mechanism that sends power to the receiver while $\xi_{max}$ represents the channel conditions.
    In the control system perspective, we treat $G(s)$ as a disturbance signal which can be modelled as either a step function, representing a sudden change in the channel conditions, or a ramp function, representing a gradual but constantly changing channel.
    
    In the case of a step function disturbance, $g(t)=g_0u(t)$, the zero located at the origin in the disturbance transfer function in \cref{eq:transfer-function-disturbance} ensures a steady state of zero, but creates an overshoot.
    %In the initial start-up, since $g>0$ to enable the plant to increase power, the overshoot is expected to be greater than $0\;dBW$ (or $1\;W$).
    The overshoot due to $g_0$ assuming $r(t)=0$ can be computed by \cref{eq:overshoot-expressions}.
    \begin{equation}
        \label{eq:overshoot-expressions}
        y_{max}=\begin{cases}
            \frac{2g_0}{eK_F} & \mathrm{critically\;damped} \\
            \frac{g_0}{K_F} & \mathrm{overdamped} \\
        \end{cases}
    \end{equation}

    In the case of a ramp function defined by $g(t)=c_0tu(t)$ where $|c_0|$ represents how fast the channel changes over time, a steady-state error can be observed defined by \cref{eq:steady-state-error-ramp}.
    This can be mitigated by increasing $K_I$ but this increase can make the system underdamped.
    \begin{equation}
        \label{eq:steady-state-error-ramp}
        y_{e}=\frac{c_0}{K_I}
    \end{equation}

     \begin{figure}[!b]
        \centering
        \includegraphics[width=\linewidth]{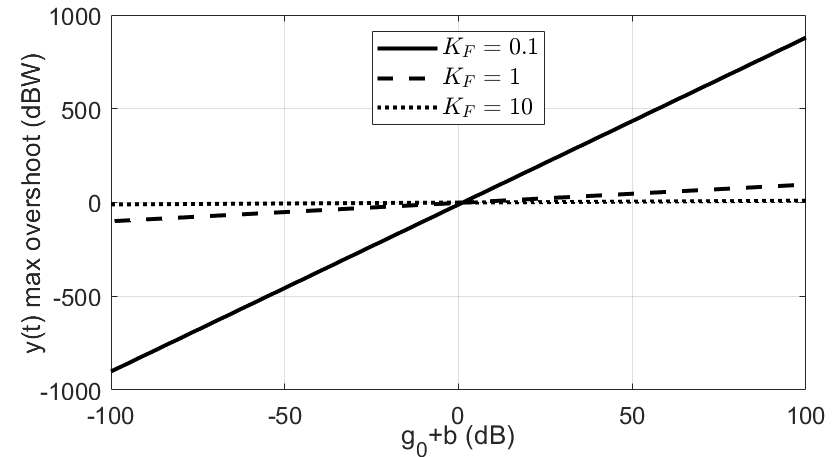}
        \caption{Maximum overshoot predicted by \cref{eq:start-up-max} with $K_I=5K_F/t_s$ for varying $K_F$ at constant $t_p=1\;ns$.}
        \label{fig:start-up-response}
    \end{figure}

    The controller can be robust to changing channel conditions as long as the change is not sudden compared to how fast the controller settles.
    It is imperative, therefore, to understand how fast the channel behaves but this is a topic for another study.

    \subsection{Start-up conditions}
    Consider the case when we want zero power at the start, i.e., $r(t)$ is a large negative number in the logarithmic domain, an initial condition defined by noise power which can range from $-160\;dBW$ to $-100\;dBW$, and a system gain $g_0>0$ that drives the system to increase power.
    In the overdamped case with the poles $p_1$ and $p_2$ (assuming $p_2>p_1$) defined from \cref{eq:poles-expression}, the start-up response of the system is shown to be \cref{eq:start-up-conditions}.
    Here, the input is $r=r_0u(t)$, the disturbance is $g(t)=g_0u(t)$, and the initial condition is $y_0$.
    
    \begin{equation}
        \label{eq:start-up-conditions}
        \begin{split}
            y(t)=(r_0+A_1e^{p_1t}+A_2e^{p_2t})u(t), \\
            A_1=-\frac{K_Ir_0+p_1g_0+p_1^2t_py_0}{p_1K_F}, \\ A_2=\frac{K_Ir_0+p_2g_0+p_2^2t_py_0}{p_2K_F}.
        \end{split}
    \end{equation}

    Since both $p_1$ and $p_2$ are negative, the transient terms $A_1$ and $A_2$ disappear over time.
    However, an overshoot at start-up can be observed from this transient response.
    The maximum value of \cref{eq:start-up-conditions} is expressed in \cref{eq:start-up-max}.
    \begin{equation}
        \label{eq:start-up-max}
        y_{max}=r_0+A_1\left(-\frac{A_1p_1}{A_2p_2}\right)^{\frac{t_pp_1}{K_F}}+A_2\left(-\frac{A_1p_1}{A_2p_2}\right)^{\frac{t_pp_2}{K_F}}
    \end{equation}
    
    To mitigate this start-up overshoot, the bias $b$ is added to the controller.
    It can be in the form of a voltage bias connected to the controlled attenuator or an initial condition forced on the controller by design.
    In effect, the initial value of $g(t)$ will be $(g_0+b)u(t)$ and we set $b<0$.
    \Cref{fig:start-up-response} shows the predicted overshoot at start-up of the controller for varying values of $g_0+b$.
    It can be observed that if $b$ is set to have a high magnitude negative value, then the overshoot can also be negative-valued.
    This condition at the output is acceptable as they can be drowned by electronic noise.

    % Over-all, the recommended parameter settings for the proposed controller design must maintain \cref{eq:settling-time} as the minimum value to avoid being underdamped and ensure that $b<-g$ to avoid a large positive-valued overshoot.

\section{Simulation}
\label{sec:simulation}
    \begin{figure*}[!b]
        \centering
        \begin{minipage}{0.3\linewidth}
            \subfloat[]{\includegraphics[width=\linewidth]{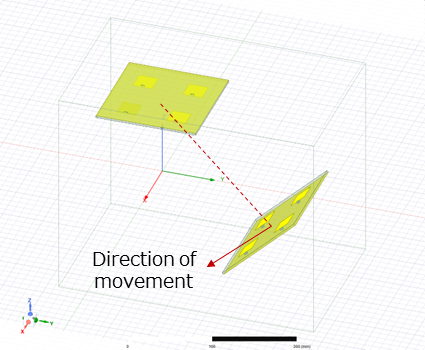}\label{fig:sim-cases1}}

            \subfloat[]{\includegraphics[width=\linewidth]{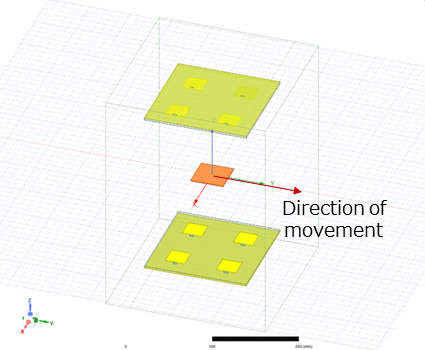}\label{fig:sim-cases3}}
        \end{minipage}
        \begin{minipage}{0.69\linewidth}
            \subfloat[]{\includegraphics[width=\linewidth]{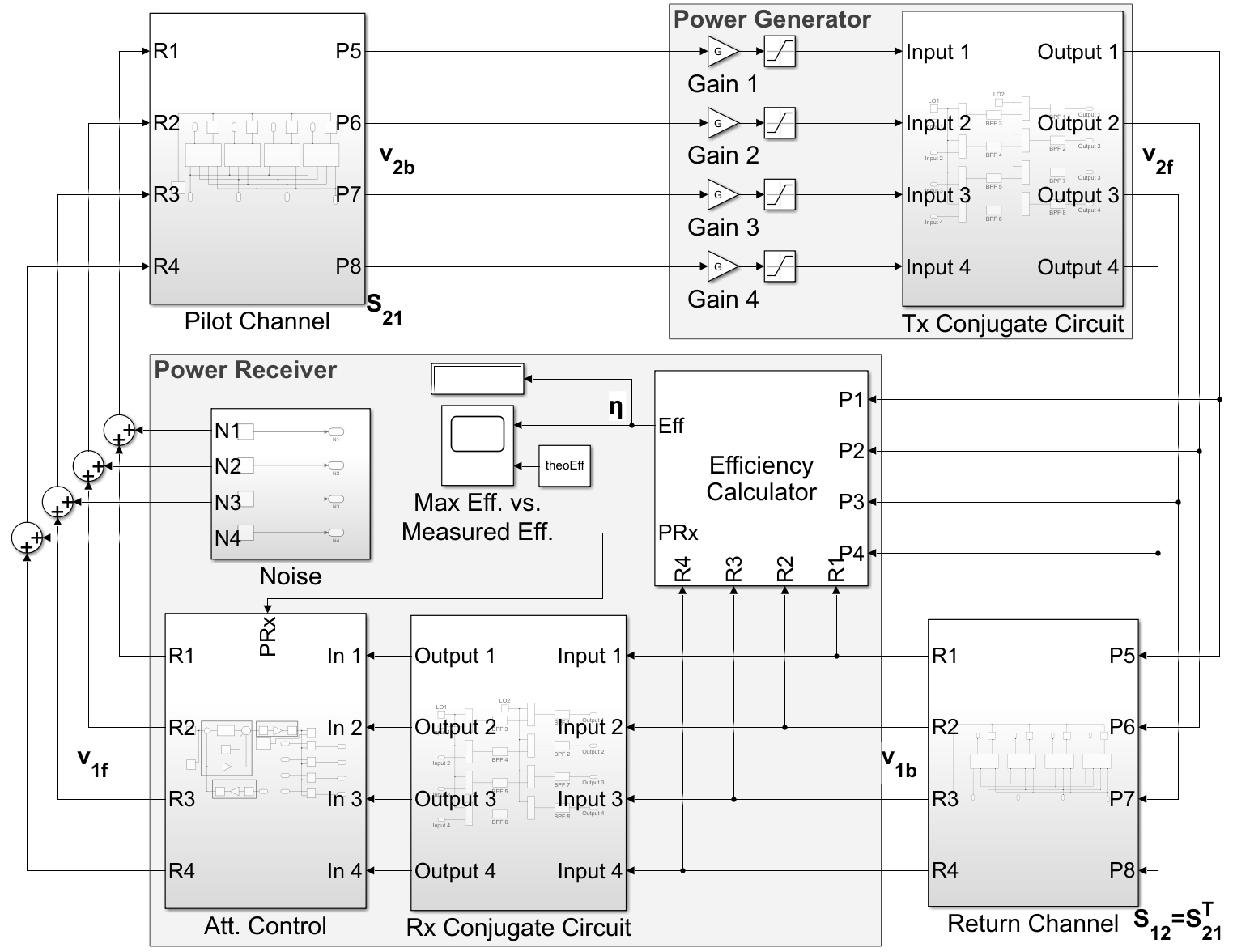}\label{fig:sim-model-total}}
        \end{minipage}
        \caption{BS-RDAA simulation model setup: (a) generator array revolves around the receiver array, (b) obstruction passes through between the arrays, and (c) MATLAB Simulink time domain model from \cref{fig:bsrdaa-block-diagram}}
        \label{fig:sim-cases}
    \end{figure*}

    \Cref{fig:sim-cases} shows the electromagnetic simulation setup created in Ansys HFSS and the time domain model created as a flowgraph in MATLAB Simulink.
    The electromagnetic simulation results were modeled into the time domain simulation to simulate a moving channel condition.

    \subsection{Electromagnetic simulation}
    Two cases were considered as portrayed in \cref{fig:sim-cases1,fig:sim-cases3} where the same 4-element generator array and 4-element receiver array are used.
    The dimension of one array is 125mm$\times$125mm or 1 wavelength by 1 wavelength, while one antenna element has a dimension of 23.7mm$\times$23.7mm.
    The spacing between them is 62.5 mm or $\lambda/2$.
    The antennas and the ground plane are copper with a 16 $\mathrm{\mu m}$ thickness, and between them is a Rogers TC600 board with a dielectric constant of 6.15 and a thickness of 2.54 mm.
    \Cref{fig:sim-cases1} is the case when the generator antenna revolving around the receiver antenna from a fixed distance of while \cref{fig:sim-cases3} is the case when an obstruction passes through the space between the arrays.
    For this simulation, this obstruction is a 50mm$\times$50mm$\times$1mm perfect electric conductor.
    Both cases place the arrays with a distance of 200 mm between them which is within the radiative near-field, but beyond the reactive near-field

    % \begin{figure*}
    %     \centering
    %     \subfloat[]{\includegraphics[width=0.32\linewidth]{images/sim-cases1.png}\label{fig:sim-cases1}}
    %     \subfloat[]{\includegraphics[width=0.32\linewidth]{images/sim-cases2.png}\label{fig:sim-cases2}}
    %     \subfloat[]{\includegraphics[width=0.32\linewidth]{images/sim-cases3.png}\label{fig:sim-cases3}}
    %     \caption{Three simulation cases used to test the performance of the controller to adjust in changing channel conditions: (a) generator array revolves around the receiver array, (b) generator array rotates on its own axis while the receiver array has a fixed orientation, and (c) obstruction passes through between the arrays.}
    %     \label{fig:sim-cases}
    % \end{figure*}

    On both test test cases, the movement of the system is assumed to be in the order of microseconds to milliseconds indicating a fast changing channel which is an extreme condition for the purpose of WPT.
    In \cref{fig:sim-cases1}, the generator antenna moves from the side towards the perfect alignment condition in 225 $\mu s$ and in \cref{fig:sim-cases3}, the obstruction moves at 800 $m/s$.
    This is done to reduce simulation time and memory usage.

    \begin{figure*}
        \centering
        \subfloat[]{\includegraphics[width=0.53\linewidth]{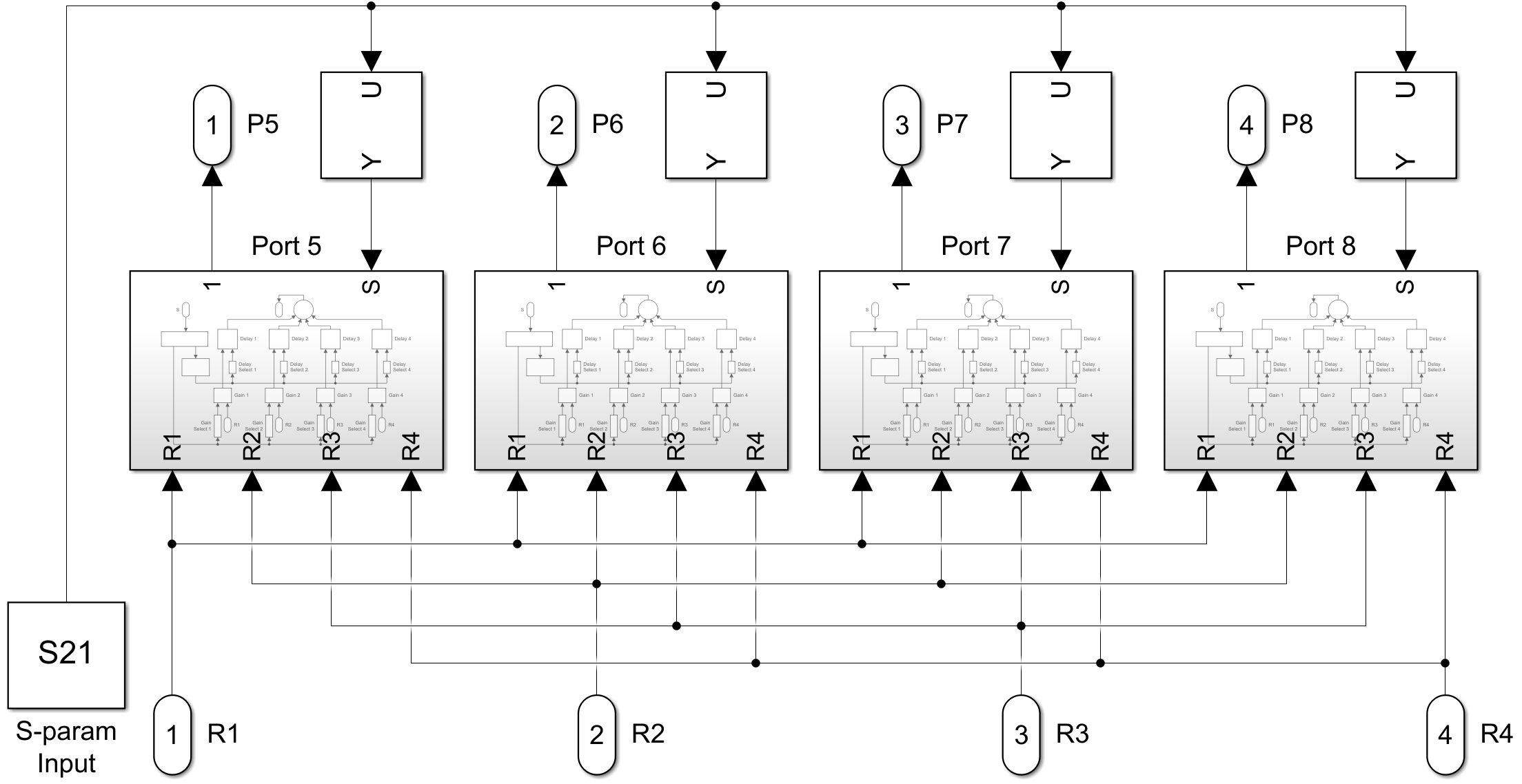}\label{fig:sim-model-s-param-a}}
        \subfloat[]{\includegraphics[width=0.46\linewidth]{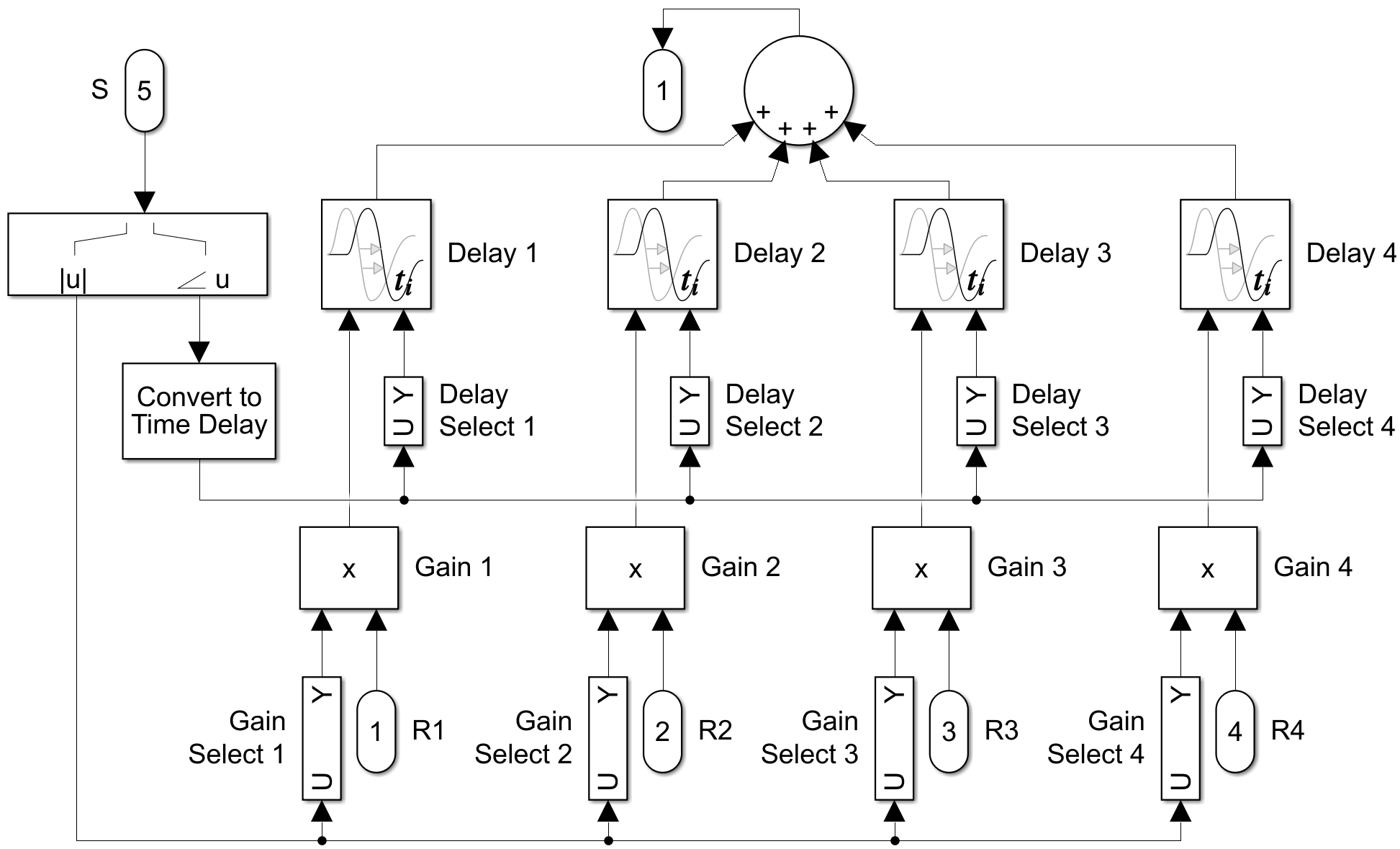}\label{fig:sim-model-s-param-b}}

        \subfloat[]{\includegraphics[width=0.34\linewidth]{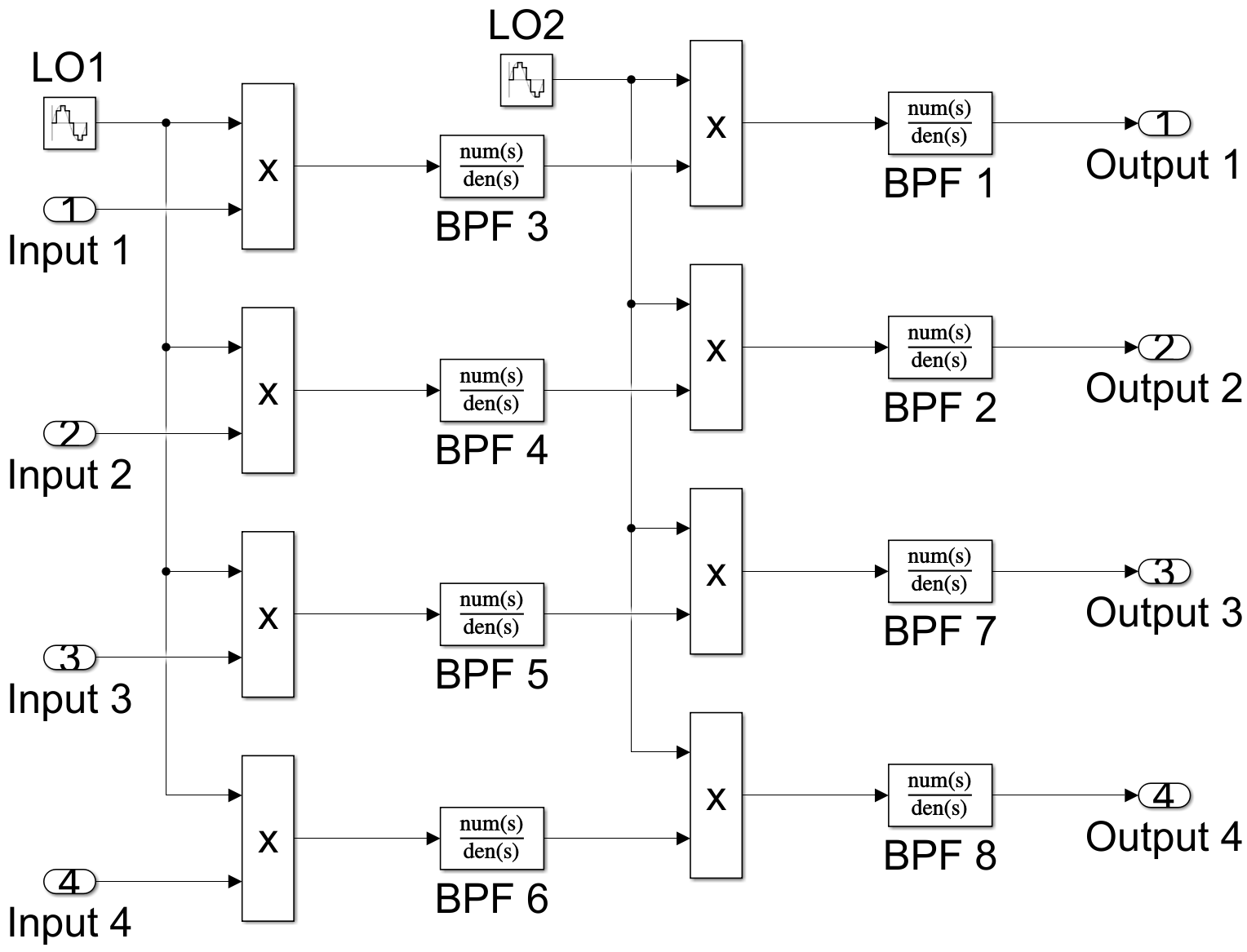}\label{fig:sim-model-conjugator}}\hfil
        \subfloat[]{\includegraphics[width=0.58\linewidth]{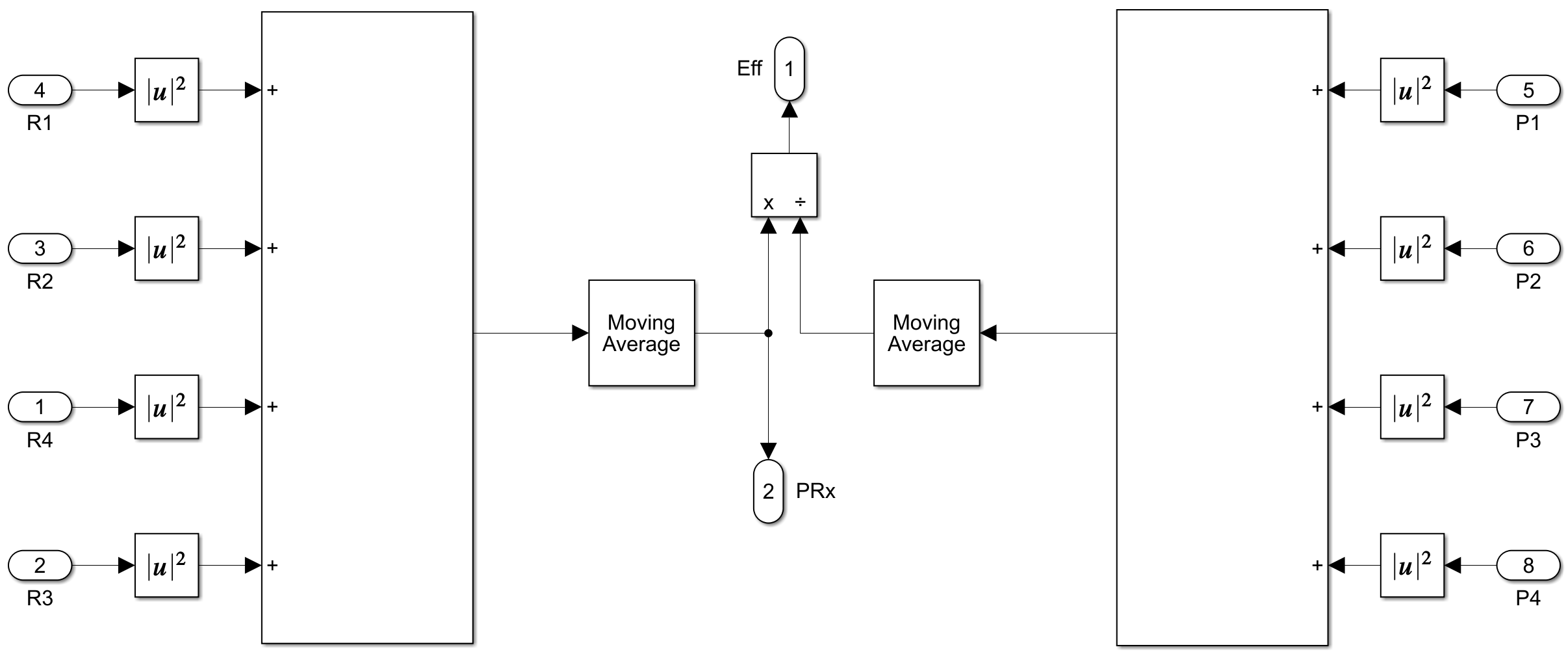}\label{fig:sim-model-eff-calc}}
        \caption{Simulink flowgraph of the different subsystems of \cref{fig:sim-model-total}: (a) S-parameter model of the whole pilot channel $\mathbf{S_{21}}$, (b) S-parameter model sending output to a specific port, (c) conjugating circuit model, and (d) efficiency calculator for data logging.}
        \label{fig:sim-model-specific}
        \vspace{-0.15cm}
    \end{figure*}
    
    \subsection{Time domain simulation model}

    The time domain model consists of the subsystems of BS-RDAA from \cref{fig:bsrdaa-block-diagram} assuming a 4-element generator array and a 4-element receiver array.
    This is visualized in \Cref{fig:sim-model-total}.
    Noise is added from the receiver side which defines the initial condition of the system.
    The pilot signal and added noise is then transmitted through the pilot channel defined by $\mathbf{S_{21}}$ after which it is boosted by the gain $G$.
    The signal was then conjugated and sent to the return channel defined by $\mathbf{S_{12}}$.
    Subsequently, the received signal is conjugated and attenuated by the controller.
    Efficiency is calculated as the ratio of the power of the output to the input of the return channel. The \texttt{ode45} continuous-time solver was chosen to remain as accurate as possible to real world conditions.
    The center frequency was set to 2.4 GHz and a maximum sampling time of 20.83 ps, or 20 samples per period, were used.
    
        \subsubsection{S-parameter model in the time domain}
        In general, the S-parameter from port $n$ to port $m$ is expressed as $S_{mn}=|S_{mn}|e^{j\phi_{mn}}$ where $j=\sqrt{-1}$.
        The contribution of the sinusoid output at port $m$ from port $n$ can be computed by $b_m=S_{mn}a_n$.
        In the time domain, if $a_n(t)=A_n\sin(2\pi f_0t)$, then $b_{mn}(t)=|S_{mn}|A_n\sin(2\pi f_0t+\phi_{mn})$ where $f_0$ is equal to 2.4 GHz in this particular simulator.
        To model the pilot channel $\mathbf{S_{21}}$, all the $N$ receiving antennas have an input that will transfer a pilot signal to port $m$ on the generator.
        Therefore, the output signal on that port is represented in the time domain by \cref{eq:s-parameter-time-domain}.
        This is also the setup for $\mathbf{S_{12}}$.
        
        \begin{equation}
            \label{eq:s-parameter-time-domain}
            b_m(t)=\sum_{n=1}^NA_n|S_{mn}|\sin(2\pi f_0t+\phi_{mn})
        \end{equation}
    
        \Cref{fig:sim-model-s-param-a} and \cref{fig:sim-model-s-param-b} show the simulator implementation of \cref{eq:s-parameter-time-domain}.
        It shows that an input matrix, $\mathbf{S_{21}}$, is fed into a subsystem model for each port output.
        This subsystem is shown in \cref{fig:sim-model-s-param-b} where the S-parameter input denoted by input 5 is broken down into its magnitude and phase components.
        The magnitude is multiplied to the input signal while the phase is converted to a time delay by $t_{mn}=\phi_{mn}/(2\pi f_0)$.
        A variable time delay block is used to realize the needed time delay.
        Finally, the sum of each contribution is connected to the output of the subsystem going back to \cref{fig:sim-model-s-param-a}.

        \subsubsection{Conjugating circuit block}
        To reflect the implementation of the experiment in \cite{AMBATALI2024experimentalvalidation}, the superheterodyne implementation \cite{miyamoto2002retrodirective, didomenico2001selfphased} of the conjugating circuit is modeled.
        The band pass filters (BPF) ensure that unwanted frequencies are mitigated.
        First, the signal is mixed with a local oscillator with a frequency that is larger than the frequency of operation.
        In this simulator, we arbitrarily chose $f_{LO1}=1.4f_0$ such that $x_{LO1}(t)=X_1\cos(2.8\pi f_0t+\phi_1)$ with arbitrary amplitude and phase $X_1$ and $\phi_1$, respectively.
        If the input is $x_0(t)=A\cos(2\pi f_0t+\phi)$ with some arbitrary phase $\phi$, the output of the band pass filter after the mixing is expressed as
        $$x_1(t)=\frac{AX_1}{2}\cos(0.8\pi f_0t+\phi_1-\phi_0).$$
    
        The second mixer uses a frequency that will upconvert the signal to the original frequency.
        In this case, $f_{LO2}=0.6f_0$, and the oscillator can be modelled by the time domain signal $x_2(t)=X_2\cos(1.2\pi f_0t+\phi_2)$.
        Similar to $x_{LO1}(t)$, $X_2$ and $\phi_2$ represents an arbitrary amplitude and phase, respectively.
        The output of the band pass filter can now be represented by $z(t)$ in \cref{eq:mixer-conjugating-output}.
        
        \begin{equation}
            \label{eq:mixer-conjugating-output}
            z(t)=A\frac{X_1X_2}{4}\cos(2\pi f_0t+\phi_1+\phi_2-\phi_0)
        \end{equation}
    
        The output of the conjugator contains evidence of phase conjugation denoted by $-\phi_0$ from the input phase $\phi_0$.
        In the simulator, we set $X_1=X_2=2$ so that no amplitude was lost in the mixer.
        The term $\phi_1+\phi_2$ represents the processing delay which is a combined effect of the filter and mixer.
        For the purpose of the simulation, a 2nd order Butterworth band pass filter with a Q-factor of $10$ was used \cite{pozar2011microwavech08, williams2014analogfilterch02}.

        \begin{figure*}
            \centering
            \includegraphics[width=0.8\linewidth]{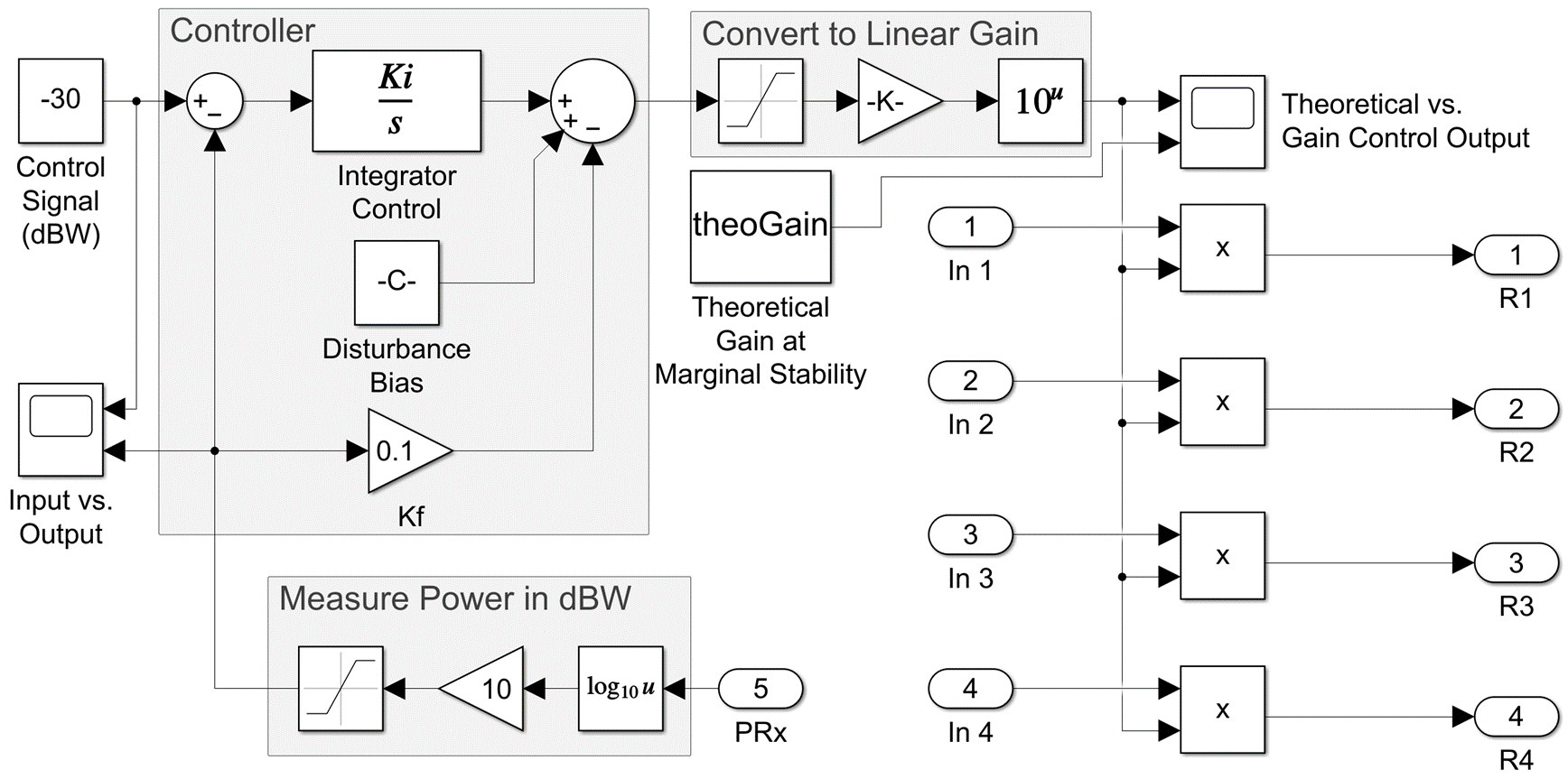}
            \caption{Simulink flowgraph implementation of the proposed controller from \cref{fig:control-block-diagram}.}
            \label{fig:sim-model-controller}
            \vspace{-0.25cm}
        \end{figure*}

        \subsubsection{Efficiency calculator block}
        The efficiency is calculated as the ratio of the power measured from the signals going out the return channel to the signals going in.
        Instantaneous power is calculated using the magnitude square of each of the ports.
        By getting the mean, the average power is calculated which reflects the power of quantities in phasor form.
        The moving average blocks in \cref{fig:sim-model-eff-calc} performs this function.
        It is set to average over 10,000 samples or approximately 0.21 $\mu$s.
    
        \subsubsection{Control system block}
        \Cref{fig:sim-model-controller} shows the MATLAB Simulink model of the proposed control design.
        To reiterate, the proposed controller design in \cref{fig:control-block-diagram} operates in the decibel measurement domain.
        Therefore, the input 5 denoting the estimate of the received power is converted to $dBW$ before being fed into the controller.
        The saturation block is added to reflect a more practical sensor which has a minimum and maximum power it is able to measure.
        Knowledge of the power in $dBW$ is used as a feedback to the control system.
        Variables \texttt{Ki}, \texttt{Kf}, and \texttt{b} define the integrator gain, feedback proportional gain, and the bias, respectively.
        Finally, the output of the controller is converted into linear gain whose saturation block also reflects a practical implementation of a controlled amplifier or attenuator which has a dynamic range.

    %\subsection{Plant model verification}

    % \begin{figure*}
    %     \centering
    %     \includegraphics[width=\linewidth]{images/startup-response-v02.png}
    %     \caption{Start-up responses of different control setups: (a) $K_F=1$, $K_I=10^7$, and $b=-120\;dB$ for $t_s=0.5\;\mu s$, (b) $K_F=1$, $K_I=10^7$, and $b=0\;dB$ for $t_s=0.5\;\mu s$, and (c) $K_F=1$, $K_I=5\times10^7$, and $b=-120\;dB$ for $t_s=0.1\;\mu s$.}
    %     \label{fig:start-up-response-sim}
    % \end{figure*}
    \begin{figure*}
        \centering
        \includegraphics[width=0.495\linewidth,height=0.25\linewidth]{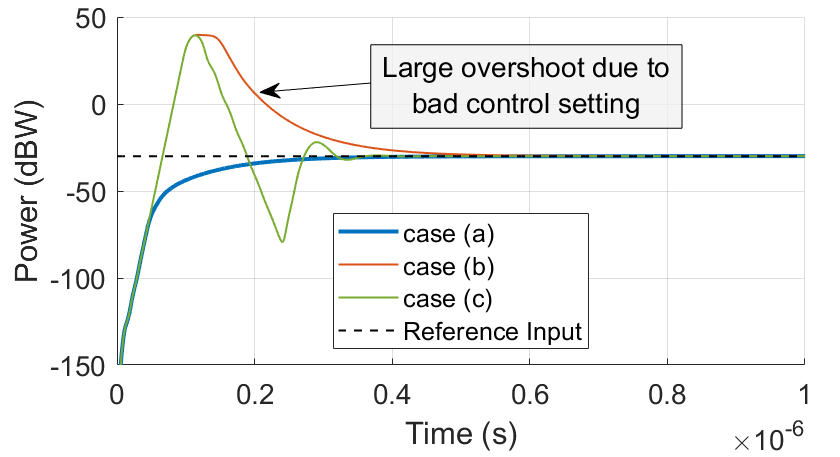}
        \includegraphics[width=0.495\linewidth,height=0.25\linewidth]{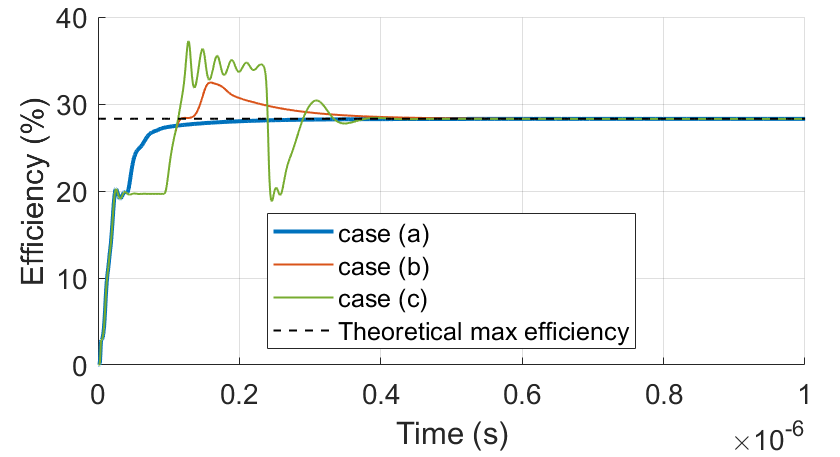}
        \caption{Start-up responses of different control setups: (a) $K_F=1$, $K_I=10^7$, and $b=-120\;dB$ for $t_s=0.5\;\mu s$, (b) $K_F=1$, $K_I=10^7$, and $b=0\;dB$ for $t_s=0.5\;\mu s$, and (c) $K_F=1$, $K_I=5\times10^7$, and $b=-120\;dB$ for $t_s=0.1\;\mu s$.}
        \label{fig:start-up-response-sim}
    % \end{figure*}

    % \begin{figure*}
        \subfloat[]{\includegraphics[width=0.495\linewidth,height=0.2\linewidth]{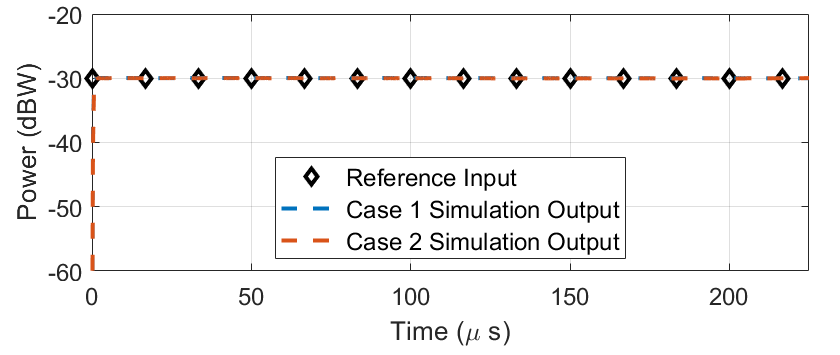}\label{fig:case-collated-a}}
        \subfloat[]{\includegraphics[width=0.495\linewidth,height=0.2\linewidth]{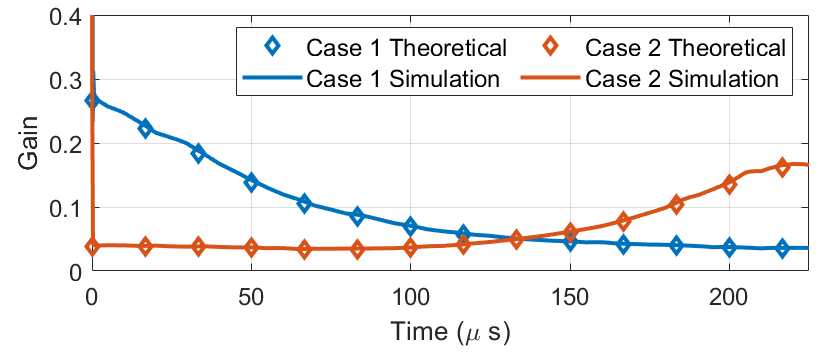}\label{fig:case-collated-b}}
        \vspace{-0.25cm}
        \subfloat[]{\includegraphics[width=0.495\linewidth,height=0.2\linewidth]{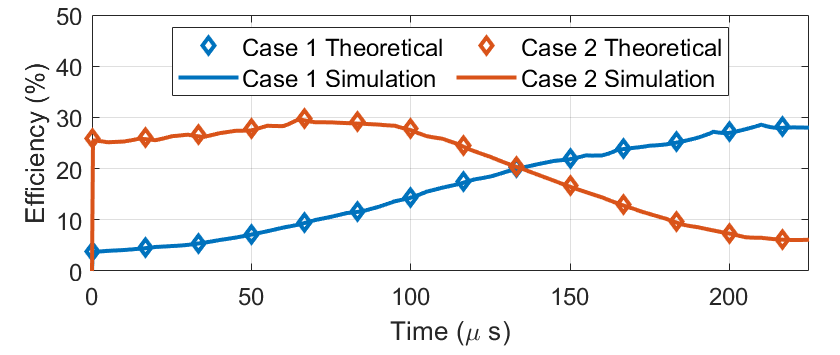}\label{fig:case-collated-c}}
        \subfloat[]{\includegraphics[width=0.495\linewidth,height=0.2\linewidth]{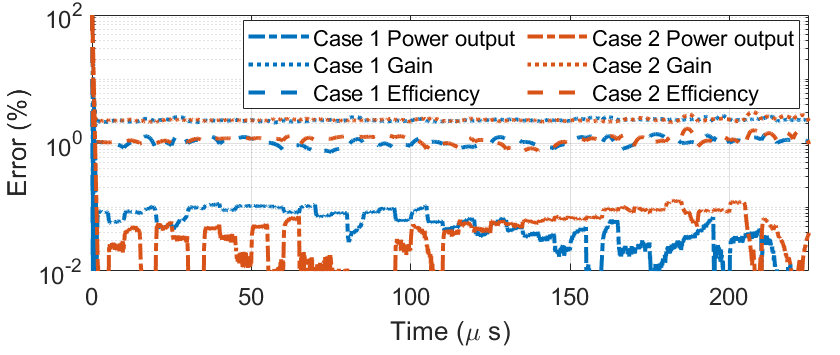}\label{fig:case-collated-d}}
        \caption{Simulation results for the two different test cases where case 1 is when the generator array revolves around the receiver array and case 2 is when obstruction passes through between the arrays.}
        \label{fig:case-collated}
        \vspace{-0.5cm}
    \end{figure*}

    \subsection{Simulation of static case}

    To demonstrate the predicted start-up response depending on the controller setup, consider the static case of \cref{fig:sim-cases1,fig:sim-cases3} where the antenna arrays are aligned with no obstruction.
    Simulations to verify the start-up responses for different cases of $K_F$, $K_I$, and $b$ are shown in \cref{fig:start-up-response-sim}.
    In case (a), a proper design is shown where $t_s>20t_p$ making it an overdamped system.
    The resulting power at start-up does not exhibit an overshoot.
    Case (b) is when no bias is applied and a significant overshoot is observed, maxing out at $40\;dBW$ which is $10^7$ times larger than the desired output power.
    Case (c) is an attempt to decrease the settling time to $0.1\;\mu s$ which is 4 times larger than the propagation time of the loop.
    This puts the system in the underdamped condition as predicted by \cref{eq:settling-time} so the overshoot and ringing in this case is as expected.

    \subsection{Simulation of dynamic case}

    To test the adaptability of the control system design, the two moving cases in \cref{fig:sim-cases1,fig:sim-cases3} are considered.
    The mean distance between the two antennas are maintained at $200\;mm$ resulting in a round trip propagation time of $1.33\;ns$.
    Adding the expected transient response of the conjugator circuit, the expected loop time is approximately $t_p=26.33\;ns$.
    The controller is now designed to have a settling time of $t_s=1\;\mu s$ around $40t_p$ which is an overdamped design.
    Letting $K_F=1$ sets $K_I=5\times10^6$.
    The bias is set to $b=-120\;dB$ to mitigate the power fluctuation at start-up.

    \Cref{fig:case-collated} shows the simulation results for the moving cases.
    The power, gain control value, and efficiency are all measured over the simulation time which ends at $225\;\mu s$.
    A transient response at start-up is observed in all cases and any overshooting was suppressed by $b$.
    The power output maintains less than $0.2\%$ error for both cases.
    This is due to the integrator feedback that removes the error from the output assuming a static channel.
    Some error is still expected due to the channel movement as predicted by \cref{eq:steady-state-error-ramp}.

    As mentioned in the theoretical analysis of BS-RDAA, maintaining its operation in marginal stability results in a steady-state that achieves maximum WPTE.
    This is demonstrated by the results in \cref{fig:case-collated} where the theoretical gain to maintain this state and the resulting efficiency is tracked by the controller.
    In terms of error, the efficiency results from both cases are within $2\%$ error while the controlled gain value is within $3\%$ of the theoretical requirement.
    Therefore, the proposed controller design is able to maintain a high efficiency WPT even if the channel condition changes as long as the settling time of the controller design can keep up with the rate of change of the channel.

    \subsection{Comparison with other control methods}

    Performance of the proposed implementation is compared with related works through simulation.
    In the position tracking case \cite{yang2023autotracking}, it is assumed that perfect estimate of the receiver position is achieved and all the transmitter elements are sending the same amplitude.
    A \textit{single-side} retrodirective antenna array was used for comparison from the pilot signal beamforming type where the pilot signal used is the transmission from a single element in the receiver array.
    For the feedback-based iterative method, the directional radiation by iterative superposition \cite{hui2020directionalradiation} was used as a reference.
    The optimization algorithm chosen for this is the Nelder-Mead method \cite{nelder-mead}.
    Two cases were simulated from this where one case is when only one element is optimized at a time while the other is where all elements are simultaneously optimized.
    One iteration is assumed to be equal to $t_p$.

    \begin{figure}
        \centering
        \includegraphics[width=\linewidth]{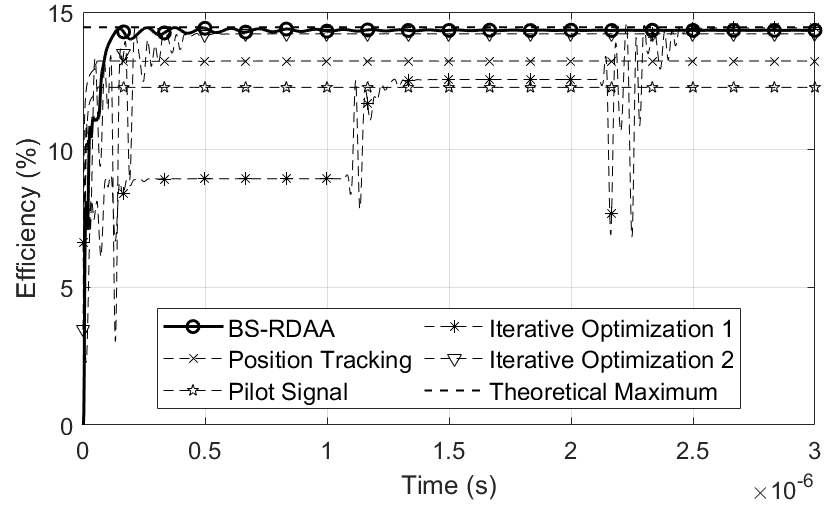}
        \caption{Transient response of different methods in the static channel case.}
        \label{fig:simulation-deg-all}
        \vspace{-0.5cm}
    \end{figure}

    The static channel case was used to compare the transient response and steady-state behavior of each control cases.
    \Cref{fig:simulation-deg-all} shows the case from \Cref{fig:sim-cases1} but the antennas are misaligned by $40^\circ$ from the normal of the transmitter antenna.
    The BS-RDAA with the proposed control system settles at 0.44 $\mu$s, the feedback-based methods at 2.56 $\mu$s for the sequential optimization case and 0.58 $\mu$s for the simultaneous optimization case, while the position tracking and pilot signal beamforming cases settle at 0.06 $\mu$s.
    It can be observed that the BS-RDAA case achieves maximum efficiency with a smaller settling time compared to other feedback-based methods, but settles slower compared to position tracking and pilot signal beamforming types.
    However, the BS-RDAA achieves maximum WPTE while the latter two types do not.

\begin{figure}
        \centering
        \subfloat[]{\includegraphics[width=0.995\linewidth,height=0.55\linewidth]{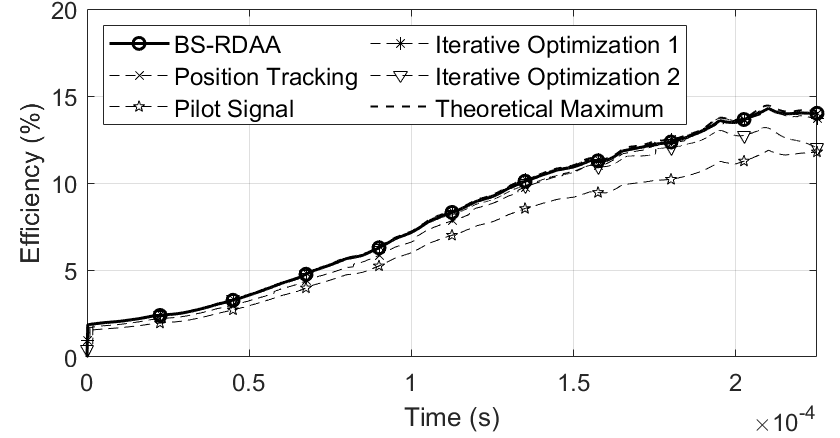}\label{fig:case1-comparison-all}}
        
        \subfloat[]{\includegraphics[width=0.995\linewidth,height=0.55\linewidth]{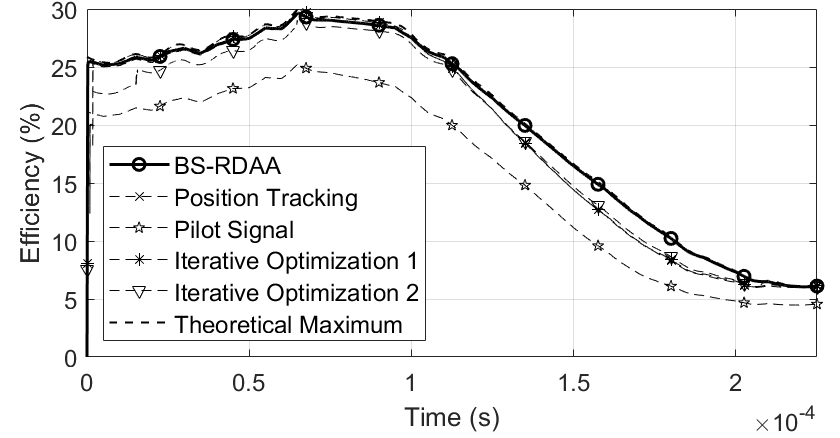}\label{fig:case3-comparison-all}}

        \subfloat[]{\includegraphics[width=0.995\linewidth,height=0.55\linewidth]{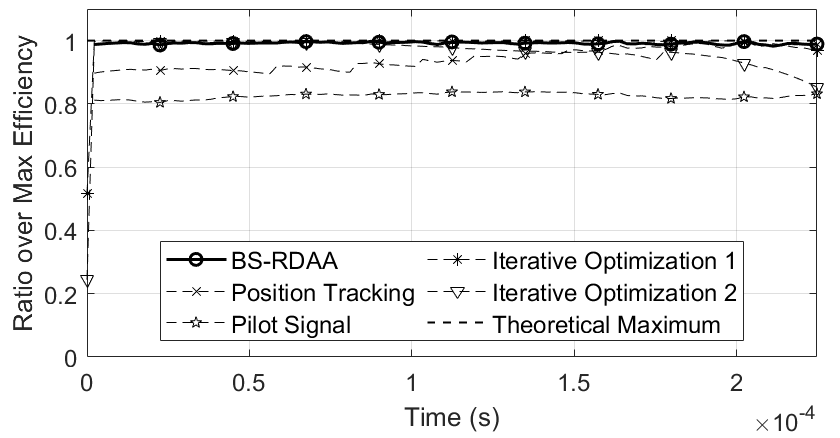}\label{fig:case1-comparison-ratio}}
        
        \subfloat[]{\includegraphics[width=0.995\linewidth,height=0.55\linewidth]{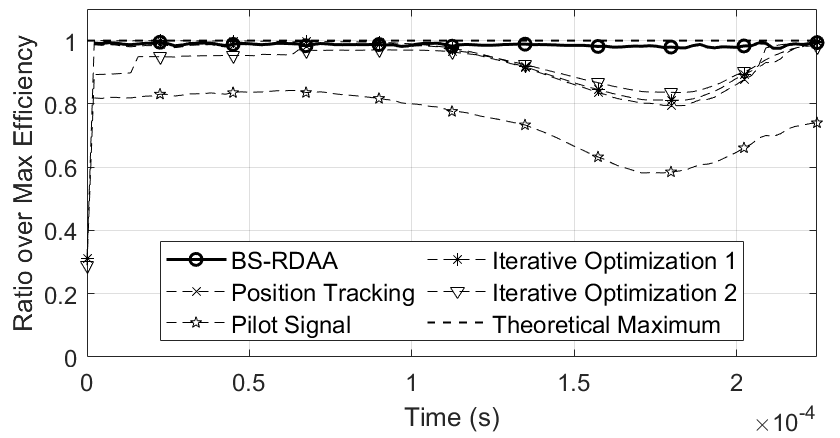}\label{fig:case3-comparison-ratio}}

        \caption{Comparison of different control methods in the dynamic channel case: (a) efficiency in \cref{fig:sim-cases1} case, (b) efficiency in \cref{fig:sim-cases3} case, (c) ratio of efficiency to the maximum efficiency in \cref{fig:sim-cases1} case, and (d) ratio of efficiency to the maximum efficiency in \cref{fig:sim-cases3} case.}
        \label{fig:case-comparison}
    \end{figure}

    The performance of the different methods were also evaluated in the dynamic channel cases illustrated in \cref{fig:sim-cases}.
    Only the efficiency was measured and compared for all cases.
    \Cref{fig:case1-comparison-all} and \ref{fig:case3-comparison-all} show the resulting efficiency and \cref{fig:case1-comparison-ratio} and \ref{fig:case3-comparison-ratio} show the ratio of the efficiency measured over the theoretical maximum.
    Even if the position tracking and pilot signal beamforming methods have a fast response, they can only achieve maximum efficiency under the condition that the arrays perfectly align with each other with no obstructions between them.
    On the other hand, the feedback-based methods can maintain maximum WPTE as long as the system does not change quickly.
    As observed in \cref{fig:case3-comparison-ratio}, both cases of iterative optimization have sub-optimal WPTE performance when the obstruction is passing through the space between the two antennas.
    Compared to all these methods, the proposed implementation can maintain its transmission close to the maximum WPTE compared to all other methods, thus, asserting its superiority in dynamic conditions.

\section{Conclusions and Recommendations}
\label{sec:conclusion}
    
    The both-sides retrodirective antenna array (BS-RDAA) system is a candidate system for long-range wireless power transfer.
    In contrast to established methods, it does not require channel estimation or iterative optimization to achieve maximum WPTE.
    However, it must always maintain marginal stability which is a time-varying condition, thus, there is a need to develop a controller for this WPT method.
    This article proposes the use of a control system to regulate the power output and stabilize the BS-RDAA wireless power transfer technique.
    A plant model view of the BS-RDAA concept was derived where the input is the loop gain and the output is the power it produces.
    Subsequently, a control system was designed based on it using techniques from Classical Control Theory, specifically, the use of a feedback gain and integrator to stabilize the system and regulate its power output.
    Performance of this proposed system was demonstrated and compared with other power beam control methods through a time domain simulation setup with the S-parameters of the system computed using an electromagnetic simulator.
    The simulation environment consisted of a static channel case, a moving antenna case, and a moving obstruction case.
    
    The error between the simulation results from theoretical expectation for the power output, efficiency, and expected gain control value are within 0.2\%, 2\%, and 3\%, respectively.
    The larger error for the latter two quantities is due to the movement of the channel which is expected to be moving slower in a practical setting.
    Therefore, the simulation results show good agreement with the theoretical analysis.
    Comparison with different works also show the superiority of the proposed system in which maximum WPTE is maintained at all times in both dynamic cases.
    Thus, we assert that the proposed control system for the BS-RDAA method is a good candidate for a long distance WPT implementation since it does not require channel estimation and iterative methods while maintaining a moderate settling time and consistently achieving maximum WPTE.

    Our direction for future work on this is to implement the experiment in hardware.
    First is to test the behavior of BS-RDAA predicted by the characteristic equation in \cref{eq:characteristic-equation} and next is to integrate a controller design.
    The current simulation can also be improved by incorporating more real-world limitations like the non-linearity of the gain/attenuation control and power measurement errors.
    Better controller design using techniques from Classical Control Theory is also a direction for future work.

% \section*{Acknowledgments}
%     The hardware experiment presented in this article is funded by Hitachi, Ltd. through their partnership with Prof. Shinichi Nakasuka and the Intelligent Space Systems Laboratory (ISSL) in The University of Tokyo.
    % Through the experiment, we were able to move from simulation and confirm our theoretical analysis in a real world implementation.

%{\appendix[Proof of the Zonklar Equations]
%Use $\backslash${\tt{appendix}} if you have a single appendix:
%Do not use $\backslash${\tt{section}} anymore after $\backslash${\tt{appendix}}, only %$\backslash${\tt{section*}}.
%If you have multiple appendixes use $\backslash${\tt{appendices}} then use %$\backslash${\tt{section}} to start each appendix.
%You must declare a $\backslash${\tt{section}} before using any $\backslash${\tt{subsection}} or %using $\backslash${\tt{label}} ($\backslash${\tt{appendices}} by itself
% starts a section numbered zero.)}


\printbibliography

% \bibliographystyle{IEEEtran}

% \bibitem{ref1}
% {\it{Mathematics Into Type}}. American Mathematical Society. [Online]. Available: https://www.ams.org/arc/styleguide/mit-2.pdf

% \bibitem{ref2}
% T. W. Chaundy, P. R. Barrett and C. Batey, {\it{The Printing of Mathematics}}. London, U.K., Oxford Univ. Press, 1954.

% \bibitem{ref3}
% F. Mittelbach and M. Goossens, {\it{The \LaTeX Companion}}, 2nd ed. Boston, MA, USA: Pearson, 2004.

% \bibitem{ref4}
% G. Gr\"atzer, {\it{More Math Into LaTeX}}, New York, NY, USA: Springer, 2007.

% \bibitem{ref5}M. Letourneau and J. W. Sharp, {\it{AMS-StyleGuide-online.pdf,}} American Mathematical Society, Providence, RI, USA, [Online]. Available: http://www.ams.org/arc/styleguide/index.html

% \bibitem{ref6}
% H. Sira-Ramirez, ``On the sliding mode control of nonlinear systems,'' \textit{Syst. Control Lett.}, vol. 19, pp. 303--312, 1992.

% \bibitem{ref7}
% A. Levant, ``Exact differentiation of signals with unbounded higher derivatives,''  in \textit{Proc. 45th IEEE Conf. Decis.
% Control}, San Diego, CA, USA, 2006, pp. 5585--5590. DOI: 10.1109/CDC.2006.377165.

% \bibitem{ref8}
% M. Fliess, C. Join, and H. Sira-Ramirez, ``Non-linear estimation is easy,'' \textit{Int. J. Model., Ident. Control}, vol. 4, no. 1, pp. 12--27, 2008.

% \bibitem{ref9}
% R. Ortega, A. Astolfi, G. Bastin, and H. Rodriguez, ``Stabilization of food-chain systems using a port-controlled Hamiltonian description,'' in \textit{Proc. Amer. Control Conf.}, Chicago, IL, USA,
% 2000, pp. 2245--2249.

% \end{thebibliography}

% \newpage

\section*{Biography Section}

\vspace{-33pt}

%\bf{If you include a photo:}\vspace{-33pt}
\begin{IEEEbiography}[{\includegraphics[width=1in,height=1.25in,clip,keepaspectratio]{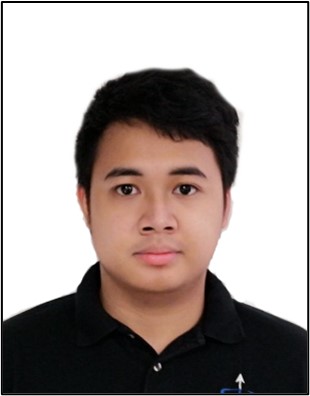}}]{Charleston Dale M. Ambatali}
(S'15-M'18) graduated from the University of the Philippines and received his BS and MS degree in 2016 and 2017, and currently a PhD candidate in The University of Tokyo expected to graduate in the spring of 2024, supported by the Japan International Cooperation Agency (JICA). From 2017 to 2018, he worked in the PHL-Microsatellite Project 1: Bus Development in which he helped develop the DIWATA-2, a 50-kg class cube satellite. From 2018, he became and is currently an assistant professor at the Electrical and Electronics Engineering Institute, University of the Philippines specializing in RF Engineering and Wireless Communications. His current PhD dissertation topic is on space-based solar power satellites looking into microwave wireless power transfer mechanisms.
\end{IEEEbiography}

\vspace{-33pt}

\begin{IEEEbiography}[{\includegraphics[width=1in,height=1.25in,clip,keepaspectratio]{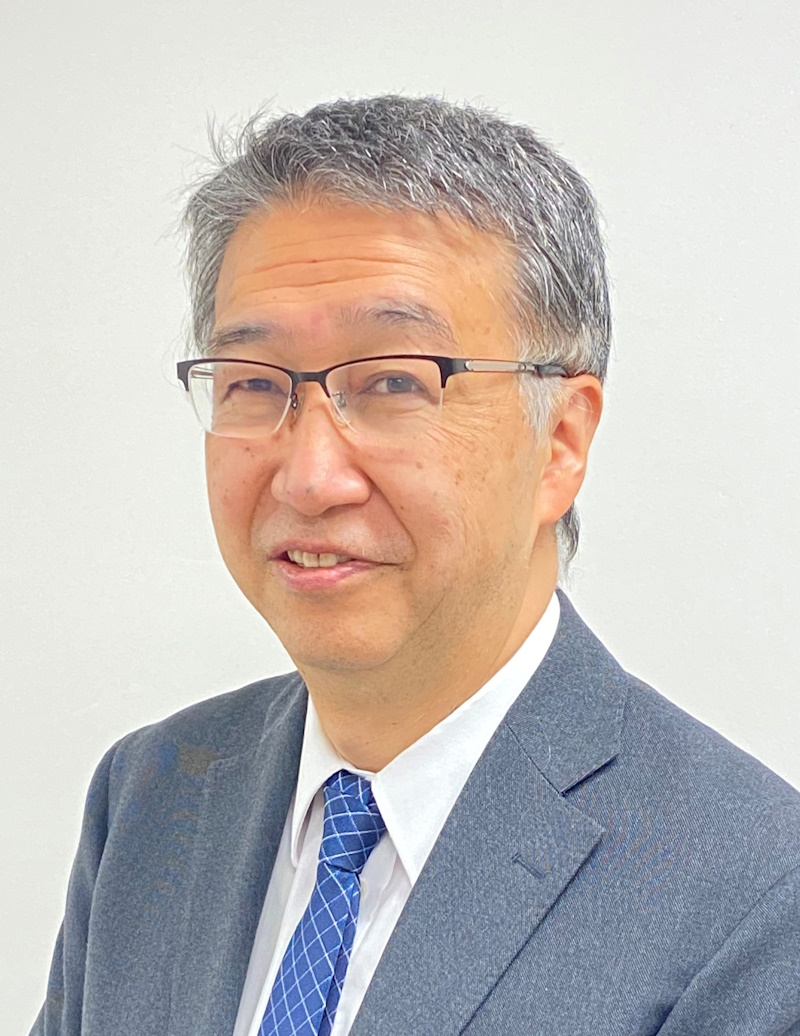}}]{Shinichi Nakasuka}
graduated from the University of Tokyo and received Ph.D in 1988.  He joined IBM Research during 1988-1990, and then has been working for Department of Aeronautics and Astronautics, University of Tokyo as a lecturer, an associate professor since 1990 and became a full professor in 2004. His major research areas include micro/nano/pico-satellites, autonomy and intelligence for space systems and guidance, navigation and control for spacecraft, and novel space systems including space-based solar power satellites. He developed and launched 15 micro/nano/pico-satellites successfully including the world first CubeSat launched in 2003.
\end{IEEEbiography}

\vfill

\end{document}